\documentclass[aps,pra,twocolumn,showpacs,groupedaddress,fleqn,floatfix]{revtex4-1}  

\usepackage{graphicx}  
\usepackage{dcolumn}   
\usepackage{bm}        
\usepackage{amssymb}   
\usepackage{amsmath}
\usepackage{epstopdf}
\usepackage[colorlinks,bookmarks=false,citecolor=blue,linkcolor=red,urlcolor=blue]{hyperref}

\begin{document}
\title{Resonantly Enhanced Tunneling and Transport of Ultracold Atoms on Tilted Optical Lattices}
\date{\today}
\author{Chester P. Rubbo}
\affiliation{JILA, (NIST and University of Colorado), and Department of Physics, University of Colorado, Boulder, Colorado 80309-0440, USA.}
\author{Salvatore R. Manmana}
\affiliation{JILA, (NIST and University of Colorado), and Department of Physics, University of Colorado, Boulder, Colorado 80309-0440, USA.}
\author{Brandon M. Peden}
\affiliation{JILA, (NIST and University of Colorado), and Department of Physics, University of Colorado, Boulder, Colorado 80309-0440, USA.}
\author{Murray J. Holland}
\affiliation{JILA, (NIST and University of Colorado), and Department of Physics, University of Colorado, Boulder, Colorado 80309-0440, USA.}
\author{Ana Maria Rey}
\affiliation{JILA, (NIST and University of Colorado), and Department of Physics, University of Colorado, Boulder, Colorado 80309-0440, USA.}
\begin{abstract}
We investigate the resonantly enhanced tunneling dynamics of ultracold bosons loaded on a tilted 1-D optical lattice, which can be used to simulate a chain of Ising spins and associated quantum phase transitions. 
The center of mass motion after a sudden tilt both at  commensurate and incommensurate fillings is obtained via analytic, time-dependent exact diagonalization and density matrix renormalization group methods (adaptive t-DMRG). 
We  identify a maximum in the amplitude of the center of mass oscillations at the quantum critical point of the effective spin system.  
For the dynamics of incommensurate systems, which cannot be mapped to a spin model, we develop an analytical approach in which the time evolution is obtained by projecting onto resonant families of small clusters. 
We compare the results of this approach at low fillings to the exact time evolution and find good agreement even at filling factors as large as 2/3. 
Using this projection onto small clusters, 
we propose a controllable transport scheme applicable in the context of Atomtronic devices on optical lattices (`slinky scheme').  
\end{abstract}

\maketitle
\section{Introduction}
\label{sec:introduction}

Ultracold atoms on optical lattices provide a well controlled system for the study of out-of-equilibrium dynamics, due to the experimental ability to tune their microscopic parameters, and to even do this  dynamically during the course of a single experiment \cite{Bloch:2008p943}.  
In recent years it has been proposed  to use these clean and tunable systems  for the realization of ultracold atom-based analogs of semiconductor electronic devices (Atomtronics)  \cite{Pepino:2009p509,Seaman:2007p2072}. 
Recent developments in single-site addressing \cite{Bakr:2009p2641,Bakr:2010p1984,Sherson:2010p2701} and in the ability to engineer optical lattices of arbitrary geometry are important  steps towards achieving this goal. 
At the heart of Atomtronics is quantum transport. 
In condensed-matter-based electronic devices the simplest way to obtain transport is to apply an electric field. 
However, for pure systems in a periodic potential, a constant force generally leads to Bloch oscillations (BO) \cite{Ashcroft:1976}.  Except from special cases such as clean semiconductor superlattices \cite{Feldmann:1992p7252}, BO are not an issue in the complex solid state environment since they are quickly damped due to strong dissipative processes, {\it e.g.}, defects, impurity scattering, and band tunneling.   Ultracold gases are on the contrary almost perfectly decoupled from the environment and therefore BO dominate the dynamics for long times \cite{Dahan:1996p4508, Battesti:2004p2768}.  These considerations pose the question as to how one may induce transport in an optical lattice while keeping full control over the atomic quantum states, as required in Atomtronics. 

Previous studies aiming to enhance transport on optical lattices have mainly focused on the regime of weak interactions in which the particles are delocalized along the system \cite{Alberti:2009p508, Gemelke:2005p2766,Ivanov:2008p2767,Sias:2007p507}. 
One reason for this is that strong interactions tend to localize particles and thus inhibit transport. 
Nevertheless, quantum transport in the strongly interacting regime has attracted some interest \cite{Micheli:2004p2862,Daley:2005p2859,Rey:2007p459,Gorshkov:2010p2866} with a focus on gaining further insights into transport of quantum information through a lattice system and quantum communication in network systems. 
Here, however, we address directly the problem of enhancing transport of particles through a lattice system by considering the strongly interacting regime of the repulsive Bose-Hubbard model (BHM) with an external linear potential. 
We show that by resonantly tuning the strength of the linear potential, {\it i.e.} by adjusting the bias between adjacent wells to allow tunneling within the lowest band, it is possible to highly control and understand, even analytically, the complicated many-body dynamics.  We assume a deep enough lattice with suppressed interband tunneling due to a large band gap.  We emphasize that what we denote as resonance should not be confused with the resonances originated by tunneling to excited Wannier-Stark ladders discussed in Ref.~\onlinecite{Gluck:2002p103}.
Based on this understanding, we propose an approach for enhancing transport, which we call the `slinky scheme' due to the peculiar nature of the resulting motion.
Somewhat counterintuitively, it is the presence of strong interactions which leads to an enhanced particle motion.    

In addition to studying BO and transport properties, strongly interacting bosons on a tilted lattice at commensurate fillings have been predicted to be a very fruitful system for the investigation of quantum magnetism in Ising models and quantum phase transitions \cite{Sachdev:2002p458}.  
The underlying idea is to map the doublon-hole excitations of a Mott insulator onto an effective spin degree of freedom when the applied tilt is tuned resonantly to the doublon interaction energy. 
Just recently the observation of the associated quantum Ising transition has been reported in the laboratory \cite{Simon:2104p2830}, leading to proposals for the realization of other systems, {\it e.g.}, quantum dimer models \cite{Sachdev:2002p458,Pielawa:2011p499}. 
These experiments also motivate the  
study of the time evolution in the resulting spin systems.  
In this paper, we do so by  
characterizing and comparing the dynamics of the center of mass (CM) oscillations of both the BHM and the effective Ising system. 
This is achieved by numerically computing the time evolution of both models using time-dependent exact diagonalization (ED) \cite{Noack:2005p75,Park:1986p2892,Hochbruck:1997p2899,Manmana:2005p63} and adaptive time dependent density matrix renormalization group (adaptive t-DMRG) \cite{white1992,white1993,Schollwock:2005p2117,Daley:2004p2943,White:2004p2941} techniques. 
The key result is that there is a maximum in the amplitude of the CM oscillation at the quantum critical point.
We extend our studies also to incommensurate fillings where the mapping to an Ising spin model does not hold, and we develop an analytical framework to deal with the dynamics in the low-density regime. 
The method relies on projecting the time evolution onto subspaces spanned by resonant families of states on small clusters. 
We compare the approximate results to t-DMRG results and find that the analytical approach provides a good approximation for filling factors as large as 2/3. 
Using this simple method one can show that the CM oscillations are restricted to have an amplitude of less than one lattice spacing.  
Even though the amplitude increases with the filling factor and is maximal at unit filling, transport remains inhibited.  
To overcome this limitation, 
we propose an alternative scheme. 
We first pattern load atoms using a superlattice \cite{Peil:2003p2891} and then stroboscopically modulate the amplitude of the lattice.
The resulting motion of the atoms is reminiscent of the motion of a `slinky' toy down a stairway. 
We expect this approach to be realizable with available experimental methods. 

The paper is organized as follows. 
In Sec.~\ref{sec:model}, we introduce the model, the non-equilibrium set-up, the mechanisms underlying the BO, and the methods used.
In Secs.~\ref{sec:commensurate_filling} and~\ref{sec:incommensurate_filling} we discuss the center of mass motion at commensurate and incommensurate fillings, respectively. 
In Sec.~\ref{sec:comparison} we discuss our numerical results obtained via time-dependent ED and adaptive t-DMRG. 
We discuss a possible signature of the critical point of the spin system in the amplitude of the CM oscillations in Sec.~\ref{sec:maximum_BO}.
In Sec.~\ref{sec:transport} we present the transport scheme mentioned above.  
In Sec.~\ref{sec:summary}, we conclude with some prospects for future work.

\section{Model, Non-Equilibrium Setup and Methods}
\label{sec:model}

\subsection{Model and Center-of-Mass observable}
We treat the one-dimensional single-band Bose Hubbard model on a tilted lattice,
\begin{equation}
\mathcal{H} = - J \sum\limits_{\langle i,j \rangle } a^{\dag}_i a^{\phantom{\dag}}_j + \frac{U}{2} \sum\limits_{i} n^{\phantom{\dag}}_i (n^{\phantom{\dag}}_i - 1) +\Omega \sum\limits_i i \, n^{\phantom{\dag}}_i,
\label{eq:bhm}
\end{equation}
where $\langle \rangle$ denotes the sum over neighboring lattice sites, $a_i^{(\dag)}$ is the annihilation (creation) operator for a boson at site $i$, and $n_i^{\phantom{\dag}} = a_i^{\dag} a_i$, and with $\Omega$ the strength of the tilting potential.  This single-band model represents a good approximation in the presence of a deep lattice which prevents inter-band transitions and decay of the atoms out of the lattice \cite{Gluck:2002p103}.
There are several ways of realizing the tilted lattice in experiments.
For example, it is possible to exploit the gravitational potential by creating a vertical optical lattice, or to detune  the counter-propagating laser-beams forming the optical lattice.
In the latter case, a time dependent detuning $\delta \mu(t)$  can  lead to an acceleration in the lattice depending on the induced velocity $v(t) = \lambda \delta \mu(t)/2$ \cite{Sias:2008p2075}, with $\lambda$ the wavelength of the laser.  
The most important observable for the treatment of the Bloch oscillations is the time evolution of the center of mass (CM) position 
\begin{equation}
x_{\rm cm}(t) = \frac{1}{N} \sum\limits_j^L  \, j \, \langle n_j(t) \rangle, 
\end{equation}
with $N$ the total number of particles on the lattice with $L$ sites, and 
$x_{\rm cm}$ is measured in units of the lattice spacing, {\it i.e.}, the lattice spacing $d$ in all subsequent calculations is set to unity. 
Measuring $x_{\rm cm}$ in the experiments \cite{Moritz:2003p2981,Fertig:2005p2980,Mun:2007p2985,Strohmaier:2007p947,Mckay:2008p2986} 
provides not only information on the CM motion and the associated current ($\dot{x}_{\rm cm}(t) = -i \langle [\hat{x},H] \rangle$, where $\hbar$ is equal to one throughout the paper) significant to transport properties, but can also provide insights into relevant energy scales in the system.  For instance, the CM motion in  Bloch oscillations has proven to be an accurate tool for metrology
\cite{Battesti:2004p2768,Ferrari:2006p2792,Carusotto:2005p2828}.  

\subsection{Bloch Oscillations}
\label{sec:BOs}

In this section, we review the main properties of Bloch oscillations (BO) of a single particle on a tight-binding chain subjected to a linear tilt.
In this case, the energy levels are discrete  $\epsilon(n) = \Omega n$ and the eigenstates are called Wannier-Stark states \cite{Wannier:1960p2940, Gluck:2002p103}. 
In the Wannier basis they are given by $|\phi_n \rangle = \sum_{j}J_{j-n}(2J/\Omega)|j\rangle$, where $J_m(x)$ is the Bessel function of the first kind.
The $n$th Wannier-Stark state is localized in the region $|j-n|<2J/\Omega$.
Due to the harmonic-oscillator like spectrum,  an initial state centered at $x_{\rm cm}(0)$ \cite{Hartmann:2004p2847} exhibits periodic CM  oscillations  with frequency $\Omega$,  
\begin{equation}
x_{\rm cm}(t) = x_{\rm cm}(0) - \frac{2 J}{\Omega} \left( 1 - \cos \Omega t \right).
\end{equation}
The same behavior is also captured by a semiclassical picture in which the linear potential is treated as a constant force dragging the particle through the Brillouin zone. 
This treatment  leads to a time-dependent value of $k$ and hence to a group velocity and to  corresponding CM oscillations of the form 
\begin{equation}\begin{split}
\dot{k} &= \Omega \Rightarrow k(t) = k_0 + \Omega t, \\
v_g(t) &= \frac{\partial E(k(t))}{\partial k} \Rightarrow x_{\rm CM}(t) \sim \frac{2J}{\Omega}\cos \Omega t,
\end{split}
\end{equation}
with $E(k)=-2J\cos(k)$ the dispersion of the system without the external potential. \\
\indent
A system of many non-interacting bosons on the lattice will show the same dynamics.
Inter-particle interactions, in contrast, tend to dampen the CM motion.  
In the weakly interacting regime, interactions lead to underdamped dynamics \cite{Trombettoni:2001p2848,Witthaut:2005p2855}, while in strong fields they can yield a multitude of interesting phenomena, {\it e.g.}, BO at interaction-induced frequencies or so-called quantum carpets \cite{Kolovsky:2003p2944,Kolovsky:2010p2945}.  
In the strongly interacting regime, the CM oscillations are almost completely suppressed.  However, time-dependent interactions can in principle stabilize them \cite{Gaul:2009p255303}.  

\begin{figure}[b]
\includegraphics[width=0.45\textwidth]{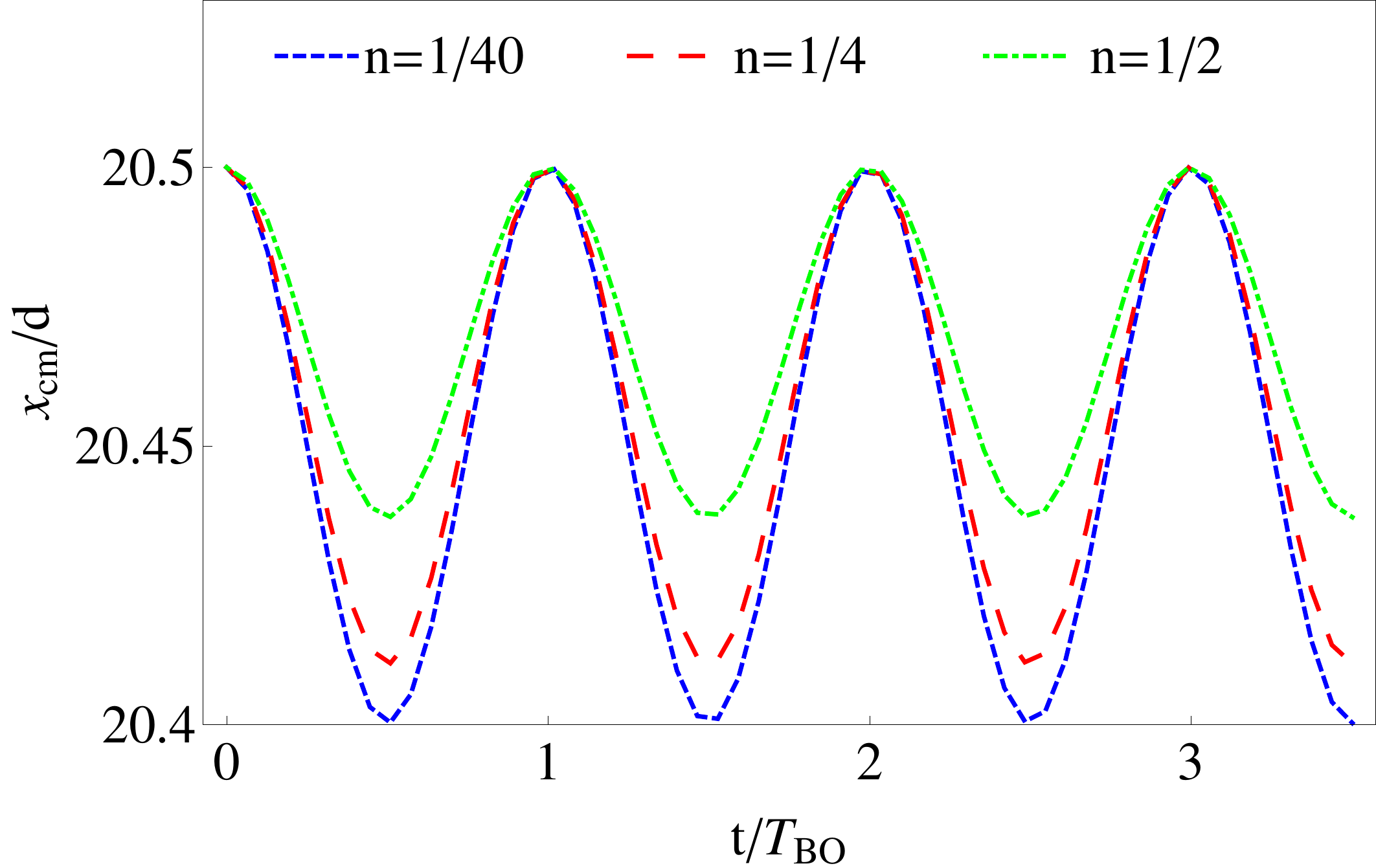}
\caption{\label{fig:fig1}(Color online) Center-of-mass motion due to Bloch oscillations for the BHM Eq.~(\ref{eq:bhm}) in the hard-core limit for different fillings $n = N/L$, where $L = 40$ sites and $\Omega = 40 J$.  The time evolution is obtained by diagonalizing the corresponding single-particle Hamiltonian and we show results for $n=1/40$ (Blue, short dashes), $n=1/4$ (Red, long dashes) and $n =1/2$ (Green, dashed-dotted line).}
\end{figure}

In the limit $U\rightarrow \infty$, a hard-core constraint is realized which mimics the Pauli exclusion principle since double occupancy is suppressed. In this fermionized regime it is then possible to introduce a Jordan-Wigner transformation,
 $a_j = c_j e^{i\pi\sum_{i=1}^{j-1}c_i^\dag c^{\phantom{\dag}}_i}$, which reduces the Hamiltonian to a tight-binding model for non-interacting
  spinless fermions. Any local observable such as the CM position is identical for spinless fermions and hard-core bosons. 
If  we denote by $|\psi^q(0)\rangle$ the initially populated single particle states with $q=1,\cdots N$,  the evolution of each of them in the basis of Wannier-Stark states reads 
 $|\psi^q(t)\rangle = \sum_n \, f^q_n \, e^{-i n \Omega t} \, |\phi_n\rangle$, with $f_n^q = \langle \phi_n |\psi^q(0) \rangle$.
Using the recurrence relation of Bessel functions (see Appendix~\ref{sec:appendixA}) for the  CM oscillations of $N$ hard-core bosons on an infinite lattice, one obtains  
\begin{equation}
x_{\rm cm}(t)  = \bar{x} + \frac{2J}{N\Omega} \, \sum\limits_{q=1}^{N} \sum\limits_{n} \text{Re}\left[e^{i \Omega  t} \, f_n^q \, f_{n+1}^{q*} \right],
\label{eq:xcm}
\end{equation}
with the average position $\bar{x} = \frac{1}{N}\sum_{q} \sum_{n} n |f_n^q|^2$. 
We find that at larger fillings the motion is suppressed, as demonstrated in Fig.~\ref{fig:fig1}. 
The simple physical picture for this effect is that as the filling increases hard-core bosons have on  average less space for free motion before they encounter each other. As the system approaches unit filling, particle transport gets fully suppressed along the lattice.
 
\subsection{Methods}

We apply different approaches for the treatment of the dynamics of the system Eq.~(\ref{eq:bhm}).
For systems small enough, we compute the time evolution by fully diagonalizing the Hamiltonian matrix.
For larger systems, we apply a Krylov-space approach to the time evolution in the framework of exact diagonalization (Krylov-ED).  For the largest system sizes treated, we apply the Krylov-space variant of the adaptive time dependent density matrix renormalization group method (adaptive t-DMRG). 
In all cases, we introduce a cutoff in the local bases on the lattice sites and keep up to 3 bosons per site.   
In the Krylov-ED approach, we approximate the time evolution operator in a basis of $m_L=10$ Lanczos vectors and use a time step of $\Delta t = 0.0005$ for the BHM.
The resulting error on the time scales treated is typically of the order of machine precision, and we can treat systems up to 15 sites with 8 particles.
With the adaptive t-DMRG, we aim for a discarded weight $< 10^{-9}$ and keep up to $m=500$ density-matrix eigenstates during the time evolution.
For the BHM, we use a time step $\Delta t = 0.0005$ and find at the end of the time evolutions displayed in the plots a discarded weight $\lesssim 10^{-7}$.
The effective spin models are much easier to treat, and we apply a time step of $\Delta t = 0.005$, resulting in a discarded weight of $<10^{-9}$ at the end of the time evolution for our largest system size of $L=50$ sites.

\section{ Center of Mass Oscillations: Commensurate Filling  }\label{sec:commensurate_filling}

In this section we study the CM oscillations at commensurate filling ($N=L$), introduce the concept of resonant dynamics and compare it to the BO.
We start the discussion with the pedagogical example of a double-well system with strong interactions $U \gg J$.

\subsection{The simplest system: a double well potential}
\label{sec:resonance_formalism}

\begin{figure}[t]
     \includegraphics[width=0.48\textwidth]{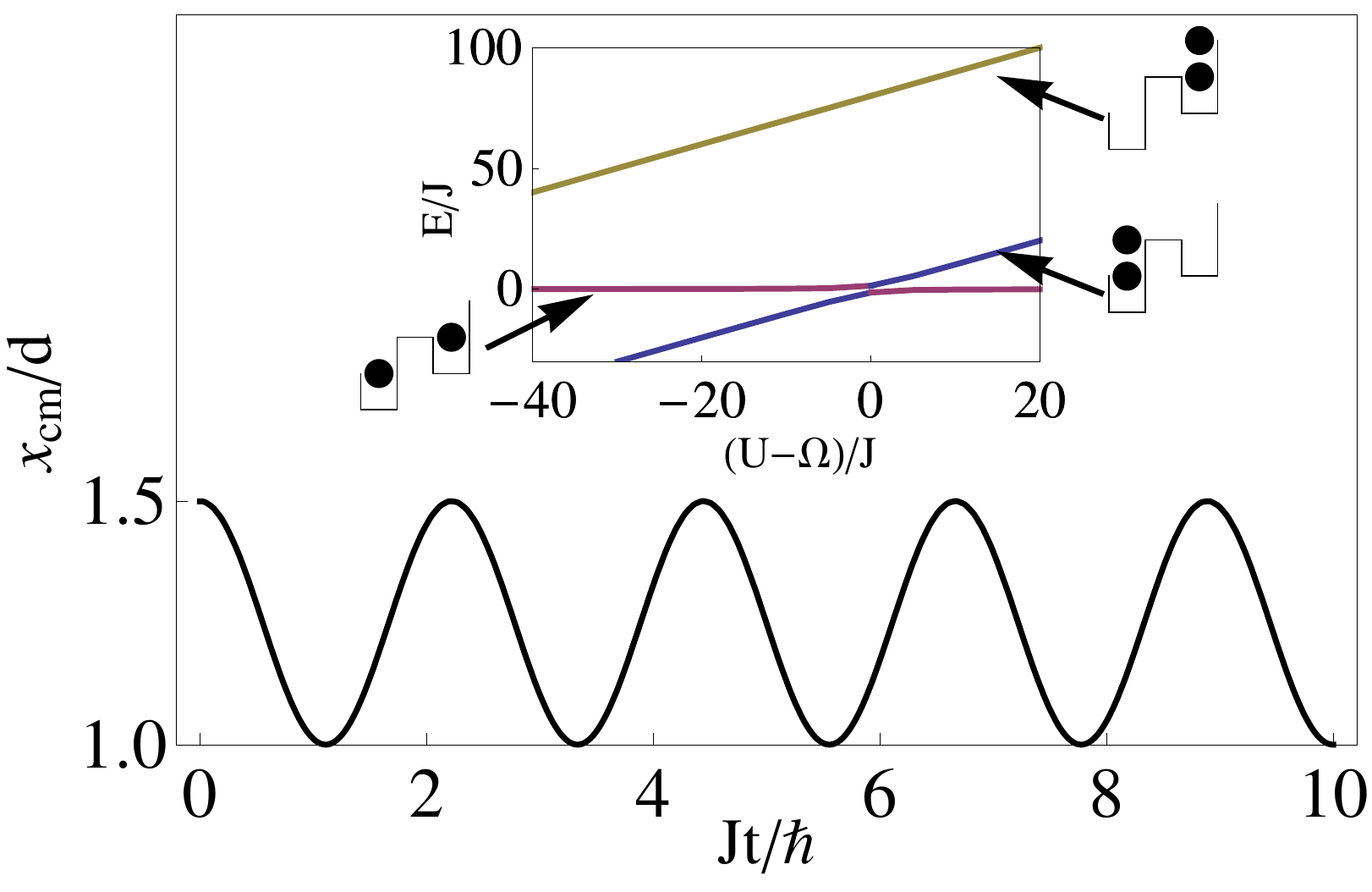}
     \caption{\label{fig:fig2}(Color online) CM motion of a double well system at resonance. 
     The inset shows the energy levels of the different states as a function of detuning $(U-\Omega)/J$. 
 } 
\end{figure}

We discuss the basic concept and properties of a {\it dipole} \cite{Sachdev:2002p458} in a tilted double well.
We assume that the system is initially in a unit-filled state $|0\rangle \equiv |11\rangle$ with one particle in each well.
In the absence of a bias, this ``Mott insulating'' state is to a good approximation the ground state for $U \gg J$.  
Excited states are obtained by moving one particle to the other well, {\it i.e.}, we obtain particle-hole excitations with two particles on one site and no particles in the other.
When applying a tilt, the lowest lying excitation is the one in which the two bosons are in the lower potential well.
This situation can be modeled by the operator $\hat{d}_i^\dag = \frac{1}{\sqrt{2}}a_i^\dag a^{\phantom{\dag}}_{i+1}$, which creates a dipole when applied to the unit-filled ground state of the untilted potential. 
For a generic  bias $\Omega<U$, the particles remain localized on each site, since the hopping can overcome the energy cost of neither the potential nor of the on-site interaction (see the inset of Fig.~\ref{fig:fig2}).
However, near resonance ($U=\Omega$), the unit-filled state and the dipole state are nearly degenerate, and so tunneling from one well to the other is possible.

The time evolution of the double-well system is given by $|\psi(t)\rangle= c_{0}(t) |11\rangle + c_{1}(t) |2 0 \rangle + c_{2}(t) |0 2\rangle$, and can be obtained analytically by an approach similar to the adiabatic elimination of a non-resonantly coupled excited state in a three-level lambda system \cite{Brion:2007p2834,Fewell:2005p2841}.  
At resonance, perturbation theory shows that the population of the off-resonant state $|0 2 \rangle$  is $\sim J^2/\Omega^2$ and hence for $\Omega  \gg J$ remains vanishingly small during the time evolution (see Appendix~\ref{sec:appendixB}).
To lowest order, the CM motion is described approximately by
\begin{align}
 &x_{\rm cm}(t)= \frac{(3 |c_0(t)|^2 + 2 |c_1(t)|^2 + 4|c_2(t)|^2)}{2}  \nonumber \\
 &\approx \frac{5}{4} + \frac{1}{4}\cos(2\sqrt{2}Jt) \nonumber \\
 &-  \frac{J^2}{8 \Omega^2}\left(1 +7\cos(2\sqrt{2}J t) + 4 \cos(\sqrt{2}J t) \cos(2  \Omega t) \right).
\label{eq:COMdoublewell}
\end{align}
The high frequency oscillation $ \sim \Omega$ is due to the  population of the non-resonant state separated by a large energy difference from the other states (Fig.~\ref{fig:fig2}).  
Since the population of this state is strongly suppressed, setting $c_2(t) = 0$ 
is a good approximation. 
Therefore, all terms $\sim J^2/\Omega^2$ are neglected, including the high frequency oscillations. 

This simple double well case illustrates that the time evolution of the system can be obtained in a good approximation by neglecting contributions from non-resonant states. 
This will be used throughout the rest of the paper. 
 

\subsection{Generic case}

\begin{figure}[b]
     \includegraphics[width=0.435\textwidth]{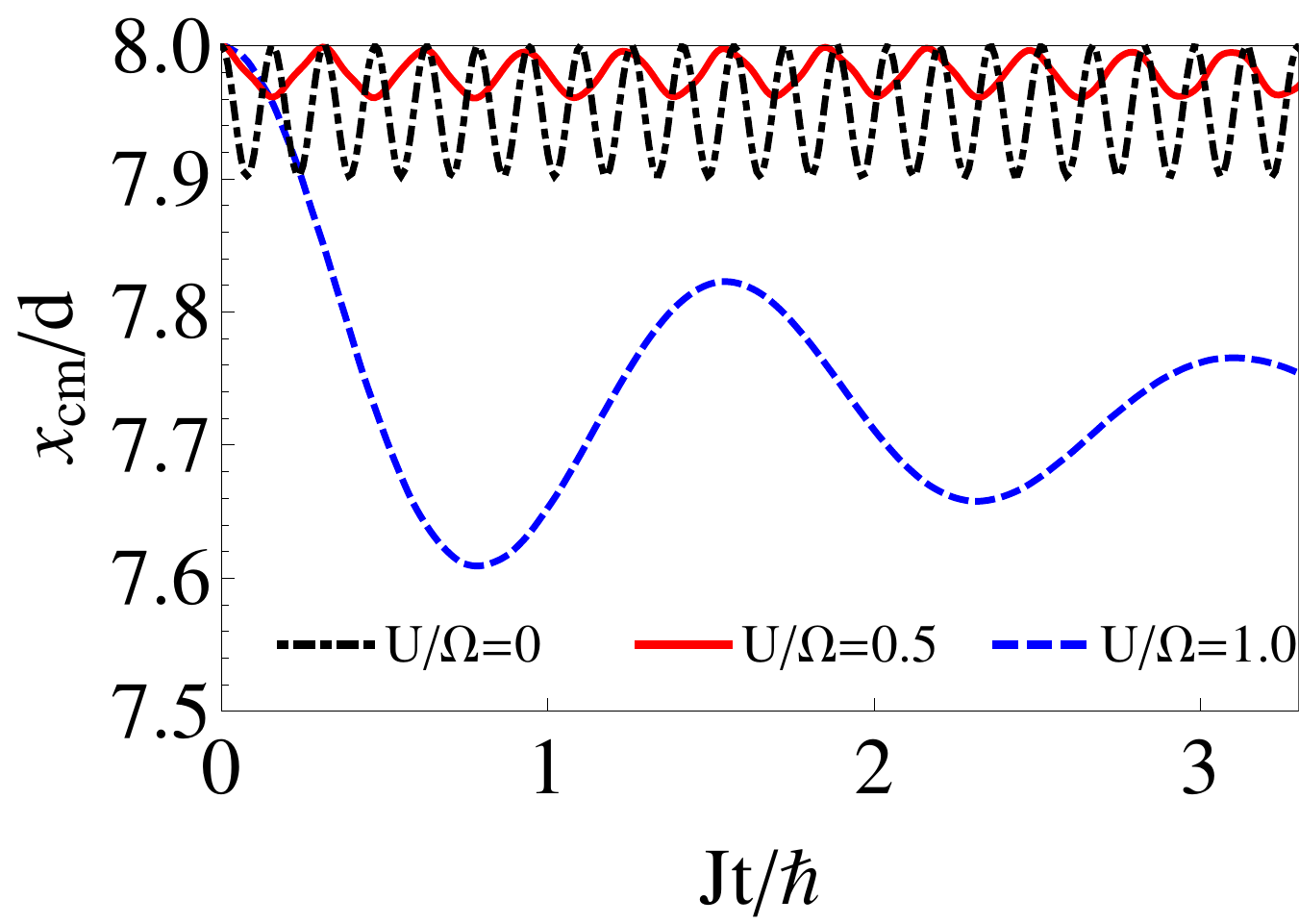}
     \caption{\label{fig:fig3}(Color online) CM motion for a BHM at commensurate filling with $L=15$ lattice sites for $J=1$, $\Omega = 40$ and three different values of $U$.  
     The results at $U = 0$ are obtained by diagonalizing the single particle problem, the results at finite $U$ are obtained by adaptive t-DMRG, with a discarded weight $<10^{-9}$ at the end of the time evolution.  
     The results for the non-interacting system show the frequency $\Omega$ and the amplitude $2J/\Omega$ of the BO, while at resonance the frequency is $\sim J$ and the amplitude gets strongly enhanced.} 
\end{figure}

We  now study the behavior of the CM oscillations as a function of $U$ while keeping $J$ and $\Omega \gg J$ fixed.  
Figure~\ref{fig:fig3} summarizes the main results. 
At $U=0$ the CM shows BO with  amplitude $4J/\Omega$ and frequency  $\Omega$. For our choice of $J$ and $\Omega\gg J$ even at $U=0$ the BO's amplitude is less than one lattice spacing. Upon increasing $U$, first the amplitude of the BO {\it decreases} by a factor of more than 2 for $U/\Omega = 0.5$, in accord with the general expectation that BO are suppressed in the presence of interactions. However, further increasing $U$ and tuning  the system to resonance, $U=\Omega$,  leads to  {\it enhanced} CM oscillations. 
The period of these oscillations scales as $J$ rather than $\Omega$, as in the case of BO, indicating that the mechanism underlying the dynamics is very different from the one of the BO 
\cite{Kolovsky:2004p477}.  
Note that even though the amplitude of the CM oscillations is maximal at resonance, it is $\leq 1/2$ lattice spacings, regardless of the system size.
This is due to the fact that the bosons are restricted to hop only between nearest
neighboring sites.  Since we analyze the resonance situation deep in the Mott-insulating regime, we require strong fields $\Omega \gg J$.  We leave out the resonance situation for weak $\Omega$ for future investigations.


\subsection{Resonant case: Effective spin model}
\label{sec:resonance}

\begin{figure}[b]
\includegraphics[width=0.48\textwidth]{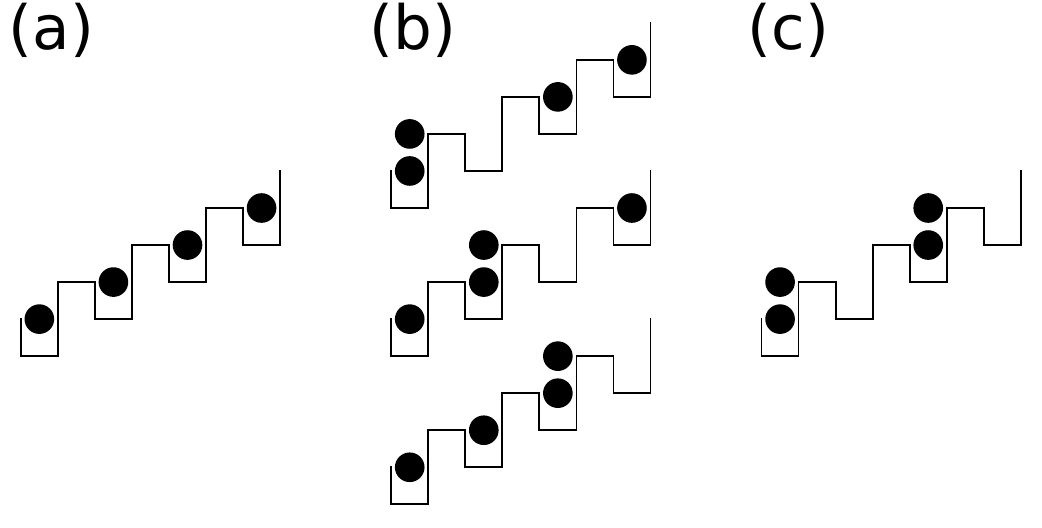}
     \caption{\label{fig:fig4} Connected resonant family of states for a system with $L=4$ lattice sites at commensurate filling. The Mott-insulating state in (a) is degenerate in energy to zeroth order in $J$ to the state configurations with one dipole excitation in (b) and the state configuration with two dipole excitations in (c).  In the spin representation, the unit filled state in (a) is represented by $|\downarrow \downarrow \downarrow \rangle$. The configurations in (b) from top to bottom are represented by $|\uparrow \downarrow \downarrow \rangle$, $|\downarrow \uparrow \downarrow \rangle$, and $|\downarrow \downarrow \uparrow \rangle$ respectively.  The configuration in (c) is represented by $|\uparrow \downarrow \uparrow \rangle$. Note that in the resonant family of states configurations with two adjacent $|\uparrow \rangle$ states are excluded.}  
\end{figure}

At resonance the dynamics is obtained by restricting to the set of dipole excitations that are degenerate in energy to zeroth order in the hopping.
We display an example for such a resonant family of states in Fig.~\ref{fig:fig4}. 
Interestingly, these states have a representation in terms of an effective spin language which we discuss below. 
In this picture, the resonant states map to all spin configurations, excluding the ones containing two adjacent $|\uparrow \rangle$ states. 

The equilibrium properties of this resonant family of states have been derived in Ref.~\onlinecite{Sachdev:2002p458}.
For the sake of clarity, we summarize them in this section.

We work in the effective basis obtained by all possible coverings of the system by dipoles, spanned by all combinations $|M;\vec{k}\rangle = \prod_{i=1}^{M}\hat{d}_{k_i}^\dag|0\rangle$, where $|0\rangle$ is the initial Mott-insulating state with one particle per site and $M$ the number of dipoles in the system; the vector $\vec{k}$ contains the positions of the $M$ dipoles on the system.
Note that two hard-core constraints appear: first, by construction, two dipoles cannot occupy the same site, leading to an on-site hard-core constraint.
Second, it is not possible to realize two dipoles on adjacent sites, since the creation of the second dipole would destroy the first one, leading to a nearest-neighbor hard-core constraint.
Taking both constraints into account, the effective Hamiltonian \cite{Sachdev:2002p458} 
\begin{equation}
H_{\rm eff, dip} = -J\sqrt{2}\sum_{i}(\hat{d}_i^\dag+\hat{d}_i)+(U-\Omega)\sum_{i}\hat{d}_i^\dag \hat{d}_i
\end{equation}
is obtained.
This model can be further mapped to a $S=1/2$ Ising model in a longitudinal and transverse external magnetic field,
\begin{equation}
\begin{split}
H_{\rm eff}&=-J\sqrt{2}\sum_{i}\sigma_{i}^{x} + (U-\Omega)\sum_{i}(\sigma_{i}^{z}+1)/2 \\
&+\Delta \sum_{i}(\sigma_{i}^{z}+1)(\sigma_{i+1}^{z}+1),
\end{split}
\label{eq:spinmodel}
\end{equation}
by introducing pseudospin raising and lowering operators via the mapping $\hat{d}_l^\dag \rightarrow \sigma_l^+$ ($\sigma_i^x$ and $\sigma_i^z$ are the corresponding Pauli matrices for the spin on site $i$).
This mapping captures the on-site hard-core constraint.
The nearest-neighbor hard-core constraint is captured by choosing $\Delta$ sufficiently large, so that two neighboring spins cannot simultaneously point up.
The quantum critical behavior of this system is governed by a single parameter, $\lambda = \frac{U-\Omega}{J}$, and possesses an Ising critical point at $\lambda_c \approx -1.85$ at which the system undergoes a phase transition from a paramagnet to an antiferromagnet, as discussed in Ref.~\onlinecite{Sachdev:2002p458}.

In this mapping, the spins are located on the bonds between two sites.
In addition, the dimension of the Hilbert space needed to obtain the time evolution via this effective model is reduced to ${\rm dim}(\mathcal{H}) = \sum_{M=0}^{\lfloor L/2 \rfloor}N^{(L)}_M = F(L+1)$, where $N^{(L)}_M =\binom{L-M}{M}$ is the number of states in a system of $L$ sites containing $M$ dipoles and $F(L)$ denotes the $L^{th}$ Fibonacci number.
This has to be contrasted with the dimension of the Hilbert space of the original model, ${\rm dim}(\mathcal{H})=\binom{N+L-1}{N}$.
Note that due to the hard-core constraints the maximum number of possible dipoles in a system is the floor function applied to half the system size, $\lfloor L/2 \rfloor$. 

\subsubsection{Dynamics in a basis of symmetric states}
One of the goals of the present paper is to identify possibly simple analytical approaches to the dynamics of the tilted Mott insulator.  In the preceding section, we have already achieved a substantial simplification by mapping the complicated bosonic system to a relatively simple effective spin model.  However, due to the hard-core constraint, the model in Eq.~(\ref{eq:spinmodel}) is a many-body model with competing interactions, so that obtaining the dynamics is still a challenging task.  In this section, we will discuss how the short time dynamics can be obtained analytically by introducing further approximations, and how the validity of this approximation can be enlarged by properly accounting for the hard-core constraint.  

The simplest approach is to completely neglect the $\Delta$-term in Eq.~(\ref{eq:spinmodel}).  This term is essentially excluding the basis states with two adjacent $|\uparrow\rangle$.  For an initial state in which all spins point downwards, for times short enough, these basis states will remain unpopulated and neglecting the hard-core constraint should be a good approximation.  This treatment should also be a better approximation for larger detunings ($|U-\Omega| \gg J$).  In this case,  dipoles become more energetically costly to populate and the system remains in the no dipole or a single dipole manifold in which the hard-core constraint  is not relevant.  This is a favorable situation: without the $\Delta$-term, we are dealing with a non-interacting system which can be treated exactly.  The dynamics in this case is obtained in a basis of Dicke states \cite{Dicke:1954p2982}, which are the set of spin states with maximum total spin.  In this approximation, the Hamiltonian can be rewritten as $H_{\rm eff}^S \approx -2 J\sqrt{2} \, \hat{S}^{x}_{\rm total} + (U-\Omega) \, \hat{S}^{z}_{\rm total}$ with collective spin operators $\hat{S}^{\alpha}_{\rm total} = \frac{1}{2} \sum_{i}\sigma_{i}^\alpha$.  The dynamics is then described as a rotation of the Bloch vector of the collective spin state manifold, $S_{\rm total} = (L-1)/2$.  It can be obtained by solving the equations of motion for the total spin components with initial condition $\langle \hat{S}^{z}_{\rm total}(0)\rangle = -(L-1)/2$.  Defining $\omega_0 \equiv \sqrt{8J^2+(U-\Omega)^2}$, one then obtains
\begin{equation}
\begin{split}
\langle \hat{S}^z_{\rm total}(t)\rangle &= -\frac{(L-1)}{2}\frac{(U-\Omega)^2 + 8 J^2 \cos{(\omega_0 t)}}{\omega_0^2},\\
\langle \hat{S}^x_{\rm total}(t)\rangle &= \frac{\sqrt{2}J (L-1) (U-\Omega)(1-\cos{\omega_0 t})}{\omega_0^2}, \\
\langle \hat{S}^y_{\rm total}(t)\rangle &=-\frac{\sqrt{2} J (L-1) \sin{(\omega_0 t)}}{\omega_0}.
\end{split} 
\end{equation}
Using the mapping discussed later in Eq.~(\ref{eq:spnmap}), the CM motion is,
\begin{equation}
\begin{split}
x_{\rm cm}(t) &= \frac{1+L^2}{2L} - \frac{1}{L}\langle \hat{S}^z_{\rm total} (t) \rangle.
\end{split}
\label{eq:8}
\end{equation}
When comparing the result of Eq.~(\ref{eq:8}) to the exact results of Fig.~\ref{fig:fig51}, we find good agreement up to times $Jt \approx 0.3$.  As discussed above, the larger the detuning from resonance, the better the qualitative agreement.  

This approximation hence leads to a closed expression for the CM motion.  However, neglecting the $\Delta$-term is a very rough approximation.  A more accurate treatment can be obtained by excluding all states with two adjacent $|\uparrow \rangle$.  In this way, we account for the hard-core constraint but stay within the collective spin manifold.  This basis we refer to as the set of symmetric states.  For a two site system (one spin), the Dicke and symmetric states are trivially the same and the time-evolved state is spanned by either set, since the hard-core constraint does not come into play.  For a three site system (two spins), the dynamics obtained in the symmetric states is exact.  However, in the general case, the hard-core constraint leads to a time-dependent phase for the different spin configurations, and the system decays out of the subspace of symmetric states.  Despite of this, for times short enough, most of the weight of the many-body wavefunction is on the symmetric manifold and one can treat the dynamics on this time scale to a good approximation.  This is done numerically.  As expected, we find that the result is a better approximation than the treament in the Dicke states, the dynamics being comparable to the exact treament for longer times.  Nevertheless, the decay out of the symmetric manifold grows, in lowest order, as $P_{\rm asym} \sim N_2^{(L)} J^6 t^6$ and therefore the time interval in which
this approximation is valid shrinks with increasing system size.

\section{Center of Mass Oscillations: Incommensurate Filling }
\label{sec:incommensurate_filling}

\begin{figure}[b]
     \includegraphics[width=0.48\textwidth]{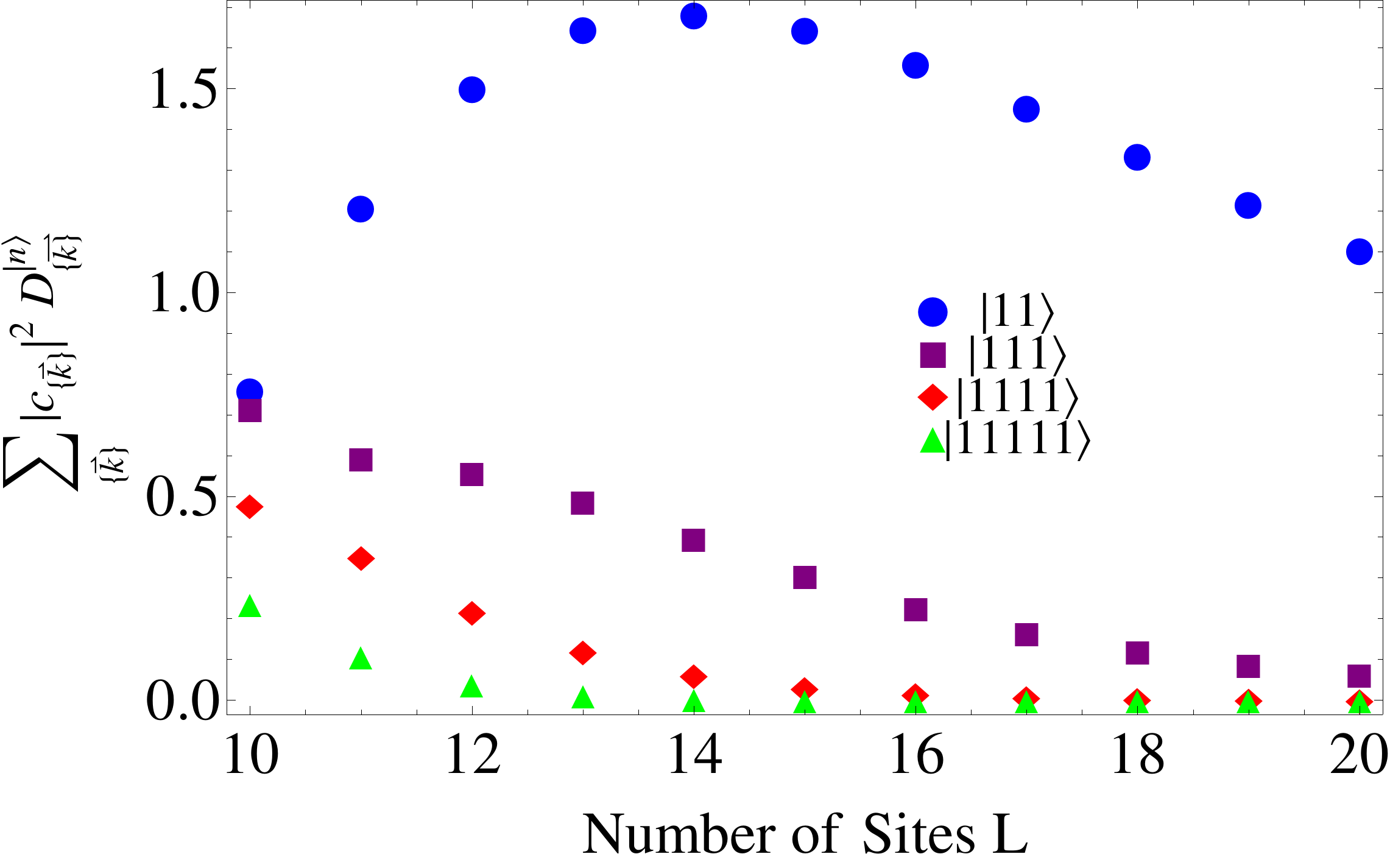}
     \caption{\label{fig:fig7}(Color online) Weighted sums of the coefficients of the ground state of the BHM, Eq.~(\ref{eq:bhm}) in the hard-core limit $U \to \infty$ for fixed $N=8$ as a function of system size $L$. 
      We consider configurations that possess $n=2$, $3$, $4$, and $5$ adjacently occupied sites (clusters), and weight the coefficients by the number of occurrences $D^{|n\rangle}_{\{\vec{k}\}}$ of the cluster which are obtained numerically as described in Appendix~\ref{sec:appendixC}. }  
\end{figure}

We  now develop  an approximate analytical solution to the dynamics of the system at incommensurate fillings, $N < L$, at which the spin model is not valid.
We begin with an illustrative example by
considering a lattice of six sites initially prepared in the Fock state $|110111\rangle$.
Suppose we time evolve this state at resonance, $U= \Omega$, with $U, \, \Omega \gg J$.
As we discussed before, the particles explore configurations degenerate in energy to the initial state in zeroth order in $J$.
The family of states which needs to be considered during the time evolution is given by the states

\begin{equation}
\begin{split}
&|\alpha_1\rangle \equiv |110111\rangle,  \\
&|\alpha_2\rangle \equiv |200111\rangle, \\
&|\alpha_3\rangle \equiv |110201\rangle, \\
&|\alpha_4\rangle \equiv |110120\rangle, \\
&|\alpha_5\rangle \equiv |200201\rangle, \\
&|\alpha_6\rangle \equiv |200120 \rangle.
\end{split}
\end{equation}
The time evolution of the system in the initial state $|\alpha_1\rangle$ is then 
\begin{equation}\begin{split}
&|\psi(t)\rangle =\\
&\cos\left(\sqrt{2}J t\right)\cos\left(2 J t\right) \, |\alpha_1\rangle+\frac{i\sin\left(\sqrt{2}J t\right)\cos\left(2 J t\right)}{\sqrt{2}} \, |\alpha_2\rangle \\
&+\frac{i\cos\left(\sqrt{2}J t\right)\sin\left(2 J t\right)}{\sqrt{2}} \, |\alpha_3\rangle +\frac{i\cos\left(\sqrt{2}J t\right)\sin\left(2 J t\right)}{\sqrt{2}} \, |\alpha_4\rangle\\
  &-\frac{\sin\left(\sqrt{2}J t\right)\sin\left(2 J t\right)}{\sqrt{2}} \, |\alpha_5\rangle-\frac{\sin\left(\sqrt{2}J t\right)\sin\left(2 J t\right)}{\sqrt{2}} \, |\alpha_6\rangle. 
\end{split}
\end{equation} 
This results in the CM motion 
\begin{equation}
x_{\rm cm}(t) = \frac{1}{10}\left[ 34 + \cos\left(2 \sqrt{2}J t\right) +\cos\left(4 J t\right)\right].
\end{equation}
The form of $x_{\rm cm}(t)$ indicates that the dynamics is governed by the two subspaces spanned by the resonant families of the $|11\rangle$ and $|111\rangle$ states. Projected onto those subspaces the effective Hamiltonian becomes
\begin{equation}
\begin{split}
\hat{H}' & = \hat{P}_D \hat{H}_{eff} \hat{P}_D = -\sqrt{2}J \left(\hat{d}_1 + \hat{d}_4 + \hat{d}_5 + h.c.\right) \\
&\equiv \hat{H}_{1} + \hat{H}_{4,5},
\end{split}
\end{equation}
where the operator $\hat{P}_D$ projects onto the resonant families.
Due to the fact that $[\hat{H}_{1}, \hat{H}_{4,5}]=0$, the time evolution takes place in  independently evolving subspaces,
\begin{equation}\begin{split}
|\alpha_1(t)\rangle &= e^{-i \hat{H}' t}|\alpha_1\rangle \\
&=e^{-i\hat{H}_1 t}e^{-i\hat{H}_{4,5}}|11\rangle \otimes |0\rangle \otimes  |111\rangle \\
&= e^{-i \hat{H}_1 t}|11\rangle \otimes|0\rangle \otimes e^{-i \hat{H}_{4,5}}|111\rangle, 
\end{split}
\end{equation} 
and the total CM dynamics reduces to the direct sum of the CM evolution in  each of the  clusters:
\begin{equation}\begin{split}
x_{\rm cm}(t) &=\frac{1}{N}\sum_j j \langle \alpha_1(t) | n_j | \alpha_1(t) \rangle \\
& = \frac{1}{N}\langle 11|e^{i \hat{H}_{1}} (n_1 + 2 n_2 ) e^{-i \hat{H}_{1}} |11 \rangle  \\
& + \frac{1}{N}\langle111|e^{i \hat{H}_{4,5}}(4 n_4 + 5 n_5 + 6 n_6 ) e^{-i \hat{H}_{4,5}} |111 \rangle \\
& =\frac{1}{10}\left[5+\cos\left(2\sqrt{2}Jt\right)+29+\cos\left(4Jt\right)\right].
\end{split}
\end{equation}
We have thus demonstrated that the resonant  dynamics of a larger system can be treated by considering the time evolution of smaller decoupled systems. In order to use  this approach for a system of arbitrary size, we have to consider configurations of particles which are sparse and spread out over the lattice.
For simplicity we will assume the atoms are prepared in the ground state in the absence of a tilt , i.e $\Omega=0$, we assume open boundary conditions and that at time $t=0$ the system is suddenly tilted close to resonance,  $U \sim \Omega \gg J$.


To be more quantitative, we  rewrite the initial wave function in a more tractable, albeit approximate way:
\begin{equation} 
\label{eq:approxgs}
|\psi_0\rangle \sim \sum_{\{\vec{k}\}}c_{\{\vec{k}\}}\prod_{\substack{ \{k_i\},\{k_j\}\\ \{k_m\}, \{k_l\} }} a_{k_l}^\dag \hat{\alpha}_{k_i}^\dag \hat{\beta}_{k_j}^\dag \hat{\gamma}_{k_m}^\dag |0\rangle,
\end{equation}
where the summation extends over all permutations of the positions $\{k_i\}$,$\{k_j\}$, $\{k_m\}$, and $\{k_l\}$; the coefficients $c_{\vec{k}}$ are derived in Appendix~\ref{sec:appendixC}. 
$|0\rangle$ is the vacuum state and the operators $a_{k_l}^\dag$, $\hat{\alpha}_{k_i}^\dag$, $\hat{\beta}_{k_j}^\dag$, and $\hat{\gamma}_{k_m}^\dag$ create configurations $|1\rangle$, $|11\rangle$, $|111\rangle$, and $|1111\rangle$, respectively, with a given weight determined by the coefficient $c_{\{\vec{k}\}}$.  We use the convention that the components of $\vec{k}$ denote the particle positions and the subscripts attached to the creation operators denote the location of the leftmost site of the cluster.  We account for subspaces up to only $|1111\rangle$ because the configurations containing larger clusters can be presumed to carry negligible weight (sparse filling condition, see Fig.~\ref{fig:fig7}).  This can be understood in the hard-core regime where the atoms in the initial many-body ground state want  to spread out symmetrically with respect to the center of the lattice in order to maximize their kinetic energy,
thus avoiding the energetically forbidden double occupancy.
 Hence, at low filling factors, the configurations that most significantly contribute to the ground state are those clusters with the lowest number of  contiguous occupied sites (see Fig.~\ref{fig:fig7}). During  the course of the time evolution, each cluster evolves independently within its  resonant manifold, allowing us to  treat the time evolution of the full system  by computing the time evolution of each of the small clusters and their associated resonant families.  In addition, the configurations $|1\rangle$ do not contribute to the dynamics since the resonance condition is significant only if at least two adjacent sites are occupied.
\\
\indent
We denote the number of $n$-particle clusters of a basis state which has particles at positions $\vec{k}$ by $D^{|n\rangle}_{\vec{k}}$.  
Then we obtain for the CM motion in this approximation
\begin{equation}
\begin{split}
& 2  N x(t) \sim \sum_{\{\vec{k}\}} |c_{\{\vec{k}\}}|^2 \left\{ D^{|11\rangle}_{\{\vec{k}\}} \cos\left(2\sqrt{2}Jt\right) \right. \\
 & \left.+ D^{|111\rangle}_{\{\vec{k}\}}\cos\left(4Jt\right) \right. \\
 &\left. +  \frac{D^{|1111\rangle}_{\{\vec{k}\}}}{34} \left[\left(1-\sqrt{17}\right)\cos\left(2\sqrt{5-4\sqrt{17}}Jt\right) \right. \right. \\
 &\left. \left.+\left(1+\sqrt{17}\right) \cos\left(2\sqrt{5+4\sqrt{17}}Jt\right)   \right. \right.\\
 &\left.\left.+ 64\cos\left(\sqrt{5+4\sqrt{17}}Jt\right) \cos\left(\sqrt{5-4\sqrt{17}}Jt\right) \right] \right\},
\end{split}
\label{eq:bigone}
\end{equation}
where we denote by $\{\vec{k}\}$ all permutations of the positions of $N$ particles.
This expression is one of the main results of the paper.
At low enough fillings, it represents a good approximation to the time evolution of systems of arbitrary size and number of particles and predicts the dynamics of tilted Mott insulators at resonance away from the commensurable case. 
In the next section, we will test the accuracy of this approach.


\section{Comparison of numerical results to the effective spin model and the analytical  treatment}
\label{sec:comparison}

In this section, we compare the time evolution obtained via the effective spin model Eq.~(\ref{eq:spinmodel}) to the exact one governed by the BHM Eq.~(\ref{eq:bhm}).
In addition, we analyze the dynamics of the effective spin model Eq.~(\ref{eq:spinmodel}) as a function of $\lambda$ and find that the amplitudes of the BO possess a maximum at $\lambda_c$.

\begin{figure}[t]
     \includegraphics[width=0.48\textwidth]{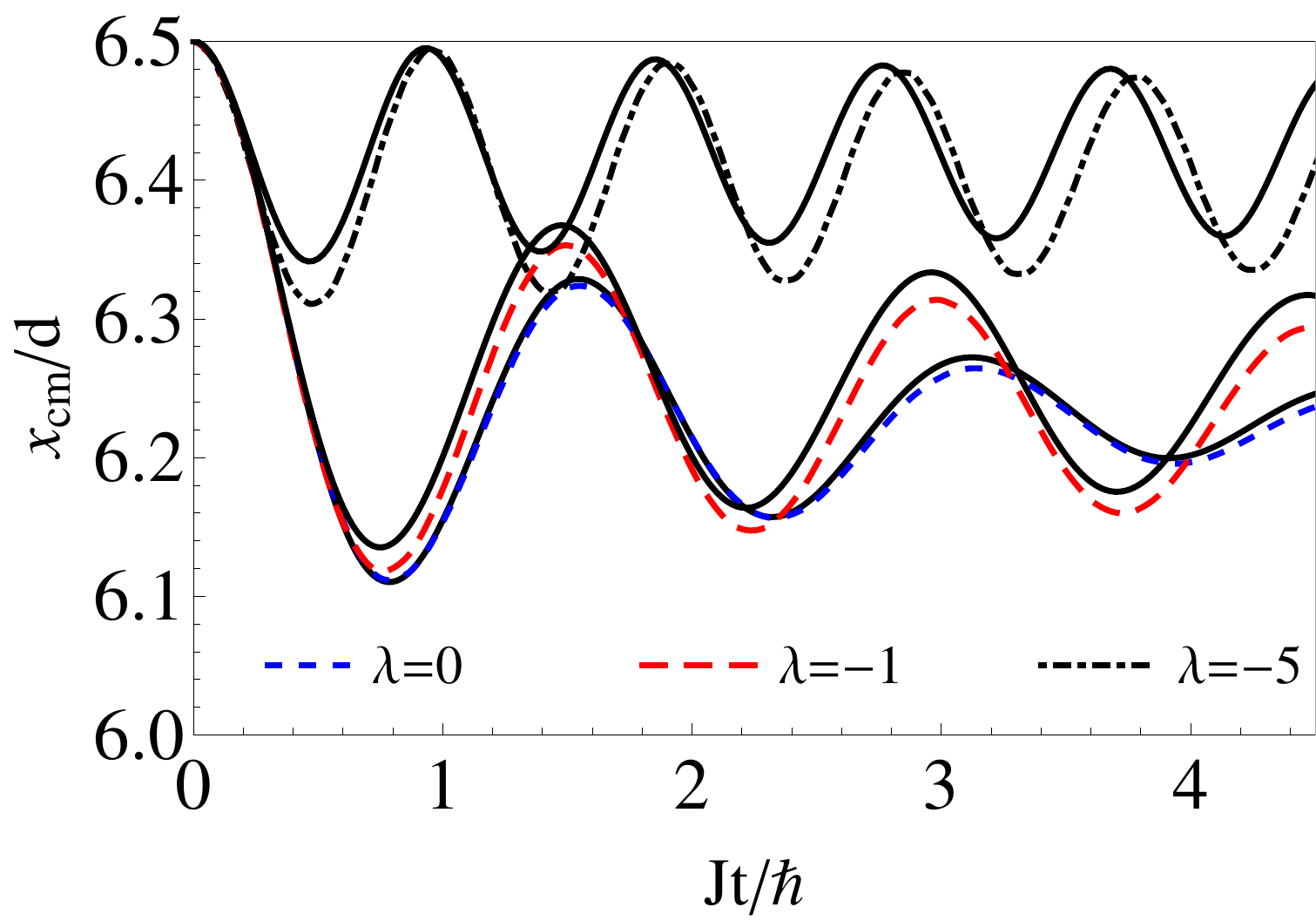}
     \caption{\label{fig:fig51}(Color online) Numerical results (Krylov-ED) for the CM motion in a system with $L =12$ sites for three different values of  $\lambda \equiv (U-\Omega)/J$, where  $J=1$ and $\Omega=40$.  
     The solid lines show the evolution of the effective spin model, while the dashed lines show  the exact evolution of the BHM. 
     As the parameters are tuned off resonance, the amplitude decreases, indicating less  participation of higher dipole excitations.} 
\end{figure}

\begin{figure}[t]
    \includegraphics[width=0.48\textwidth]{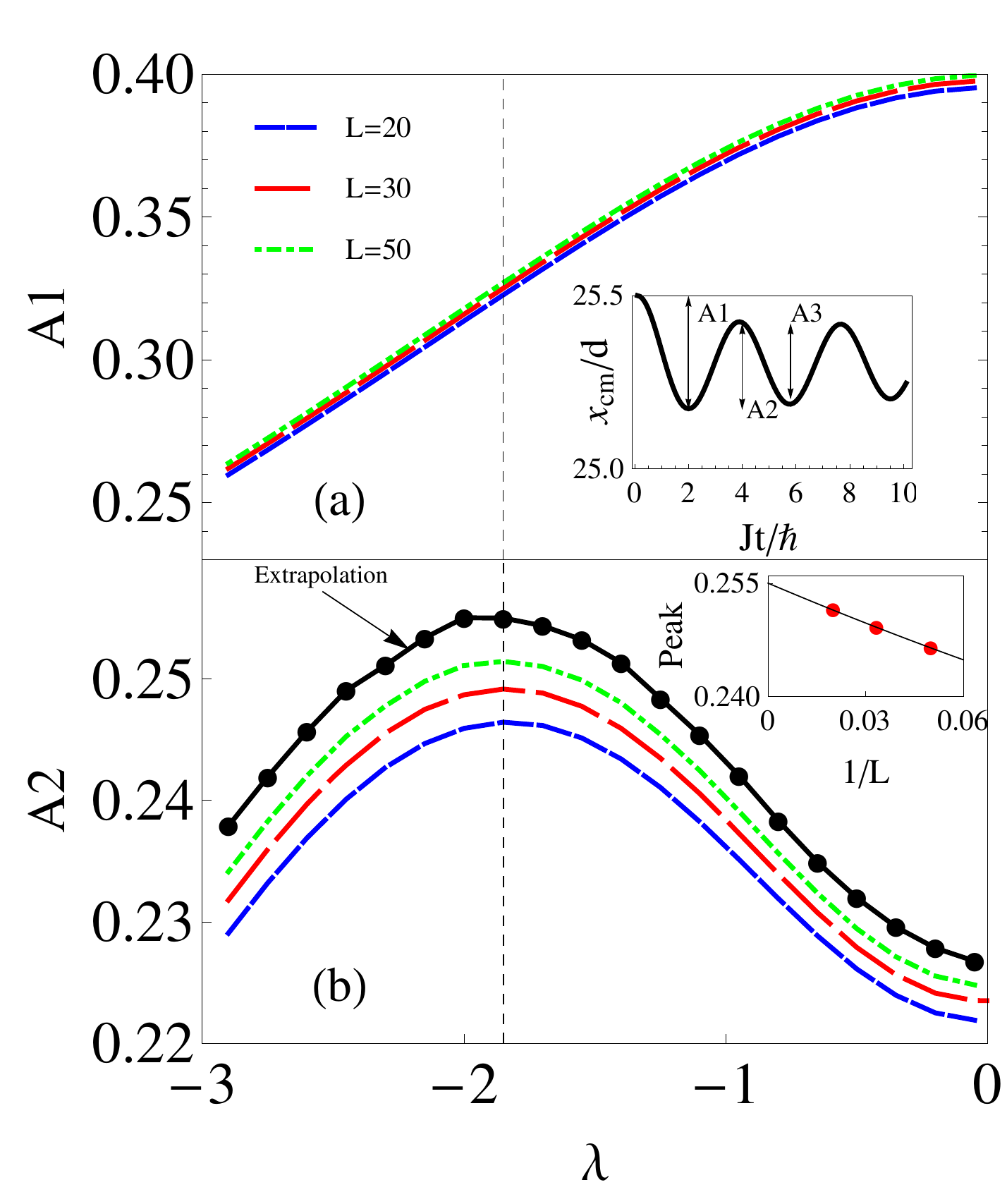}
\caption{\label{fig:fig41} (Color online) Amplitudes of the CM motion as obtained in the effective spin model and as defined in the inset in (a), which shows the evolution at resonance. The results shown are for systems with $L=20, \, 30$ and 50 lattice sites and are obtained using the adaptive t-DMRG.  The amplitudes are obtained from interpolating the discrete data points of the CM motion.  (a) Difference $A1$ (measured in units of lattice spacings) between the initial CM position and the first minimum as a function of $\lambda$.  
 (b) The second amplitude $A2$ (measured in units of lattice spacings) as a function of $\lambda$.  
 The inset shows the finite size extrapolation at $\lambda = -1.85$, and the black circles show $A2(\lambda)$ after finite size extrapolation.  We estimate the error of the extrapolation to be of the order of the symbol size.  
 } 
\end{figure}

\begin{figure}[t]
\includegraphics[width=0.48\textwidth]{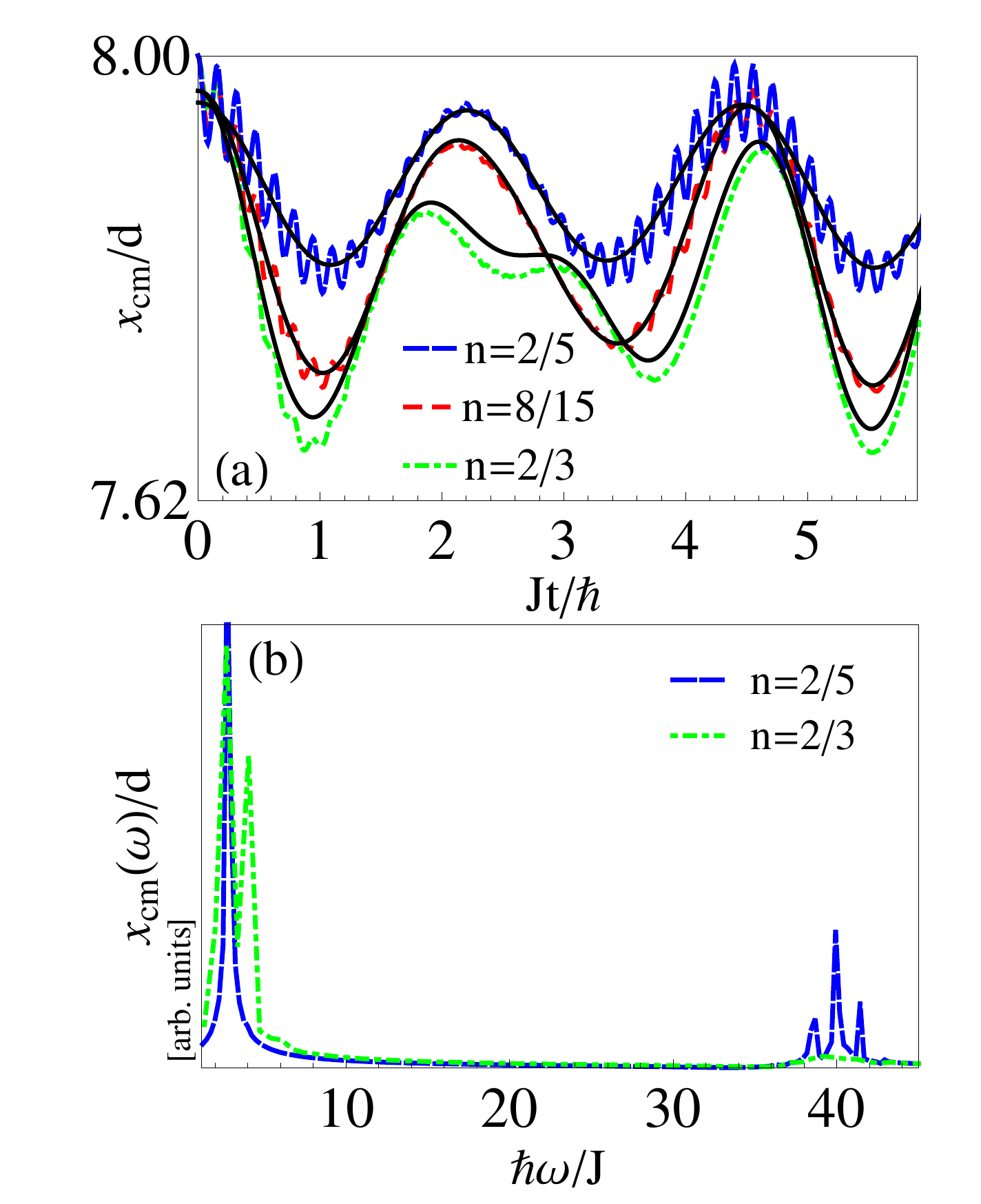}
\caption{\label{fig:fig8} (Color online)  (a) CM motion for a system of $L=15$ lattice sites at resonance at low fillings away from commensurability.  
The solid curves show the result of the analytic calculation, Eq.~(\ref{eq:bigone}).   
Note that the analytic expression neglects the fast BO which are caused by population of non resonant states.  
(b) Spectral analysis of the CM motion.  At low fillings, $2$ particle cluster states dominate the  dynamics.  The high frequency BO are visible as peaks at $\Omega/J = 40$ which decrease upon increasing the filling. The additional low frequency peaks are due to larger cluster states which become relevant at higher fillings.
The results for the BHM at $n = 2/5$ and $n = 8/15$ are obtained via Krylov-ED, the ones at $n=2/3$ are adaptive t-DMRG results. 
}    
\end{figure}

\subsection{Accuracy of the effective spin dynamics at commensurate filling}
\label{sec:accuracy_spindynamics}

In order to compare the dynamics of the effective spin model to the one of the BHM at  commensurate filling, we use the mapping
\begin{equation}
\begin{split}
\text{left boundary site:} \quad n_1 &\rightarrow (\hat{\sigma}^z_1+3)/2\\
\text{right boundary site:} \quad n_{L^\prime} &\rightarrow (1-\hat{\sigma}^z_{L})/2\\
\text{bulk:} \quad n_i  &\rightarrow (\hat{\sigma}^z_i-\hat{\sigma}^z_{i-1}+2)/2
\end{split}
\label{eq:spnmap}
\end{equation}
where $L$ is the size of the spin system and $L^\prime = L+1$ the size of the bosonic system, which has one lattice site more than the spin system, due to the fact that the dipoles are located on the bonds.
This mapping is obtained by comparing individual site occupations in the presence or absence of a dipole; the different mapping at the boundary sites and in the bulk is due to the open boundary conditions (OBC).
The difference between the mapping at the left and the right boundaries is due to the fact that the site at the highest potential (right boundary) will
 possess either zero or one particle, but the site at the lowest potential (left boundary) has either one or two particles.

The dynamics of the BHM is obtained by evolving an initial Mott insulating ground state with one particle per site.
In the spin picture, this is equivalent to an initial state with all spins pointing downwards.
In Fig.~\ref{fig:fig51} we show the time evolution of the BHM and the one of the effective spin model at $\lambda = 0$, $\lambda = -1$, and $\lambda = -5$. 
We find that the effective spin model reproduces the dynamics of the BHM even when detuning not too far off resonance.
The evolution of the systems is essentially identical on short time scales ($Jt \lesssim 2.5$ for $\lambda = 0$, $Jt \lesssim 0.5$ for $\lambda = -1$, and $Jt \lesssim 0.3$ for $\lambda = -5$) but then differs increasingly  in the course of the time evolution.
In addition, for small detunings from resonance, the frequency of the oscillation is in excellent agreement to the exact solution, while it differs in the case of stronger detunings.
Note that detuning causes the system to be rigid in the sense that states that contain more dipoles are more costly in energy.
Hence, on sufficiently short times, the number of dipole states contributing to the dynamics is reduced so that the time evolution can be obtained by considering a smaller number of states (see Ref.~\onlinecite{brandon_thesis} for a detailed discussion). 

\subsection{Maximum of the CM amplitude at $\lambda_c$}
\label{sec:maximum_BO}

Now we turn to the properties of the CM oscillations in the time evolution of the effective spin system Eq.~(\ref{eq:spinmodel}) upon changing $\lambda$.
As before, we consider the time evolution of an initial state in which all spins are pointing downwards, which is equivalent to a Mott-insulator with one particle per site in the bosonic language.
As shown in Ref.~\onlinecite{Sachdev:2002p458}, the effective spin model possesses a quantum critical point at $\lambda_c \approx -1.85$. 
We are interested in possible signatures of this critical point in the CM motion. 
In Fig.~\ref{fig:fig41} we present our adaptive t-DMRG results for the first oscillations $A1$, $A2$ and $A3$ [defined in the inset of Fig.~\ref{fig:fig41}(a)] as a function of detuning $\lambda$ for systems with $L = 20, \, 30$ and 50 sites.
Interestingly, we find a local maximum of $A2$ at $\lambda \approx -1.85$, indicating that this quantity might indeed reveal the existence of a critical point in the time evolution of this system.
This feature seems to persist upon changing the system size, and at $\lambda_c$ we obtain after  finite size extrapolation the value $A2(\lambda_c,L\to\infty) \approx 0.255$.
This can be contrasted to the value of $A2(-0.05,L\to\infty) \approx 0.227$, so that the amplitude when increasing the detuning from resonance at $\lambda=0$ to $\lambda_c$ changes by $\approx10\%$. 
Similarly, we find a maximum of $A3$ at $\lambda \approx -1.85$.
However, such a maximum in the vicinity of $\lambda_c$ does not show up in $A1(\lambda)$, a quantity which instead reaches its peak value at resonance $\lambda = 0$. 
\\
\indent 
This puts forth the interesting possibility that one may be able to use the CM oscillations to identify quantum critical points of a generic phase transition. 

\subsection{Accuracy of the effective cluster dynamics at incommensurate fillings}
\label{sec:accuracy_clusterdynamics}

In Fig.~\ref{fig:fig8} we compare the CM motion as obtained by the original BHM for a system of $L=15$ sites at various fillings $n \leq 2/3$ to the one obtained in the approximate treatment using families of small cluster states.
At fillings $n=2/5$ and $n=8/15$ the plot shows data obtained via the Krylov-ED approach, so that the results are essentially exact.
At $n=2/3$, we present data obtained via adaptive t-DMRG where the discarded weight is $>10^{-9}$ for times $J t > 2$, so that the results at later times might be affected by numerical errors larger than the width of the lines shown in the graph.
The overall behavior of the approximate solution given by Eq.~(\ref{eq:bigone}) is in good agreement with the numerical results in all cases shown.
However, at low fillings, fast oscillations with frequency equal to $\Omega$ [Fig.~\ref{fig:fig8}(b)] are superimposed onto the resonant frequency oscillations.
These high frequencies are due to the single-particle BO and are not taken into account by our projection onto resonant families of states. 
However, these BO can be suppressed by increasing the filling [as seen in Fig.~\ref{fig:fig8}(b)] or by using larger values of $\Omega$ while ensuring the resonant condition.
This is explained by the fact that the amplitude of the BO is $\propto 1/\Omega$.
At larger fillings, the approximation Eq.~(\ref{eq:bigone}) breaks down since clusters larger than the $|1111\rangle$ states become relevant.\\ 

With this, we conclude our treatment of the dynamics after a sudden tilt of an initial Mott-insulating state and turn now to the question of how to enhance transport in these systems. 

\section{Engineering Transport: A Slinky Scheme}
\label{sec:transport}

\begin{figure}[h!]
\includegraphics[width=0.38\textwidth]{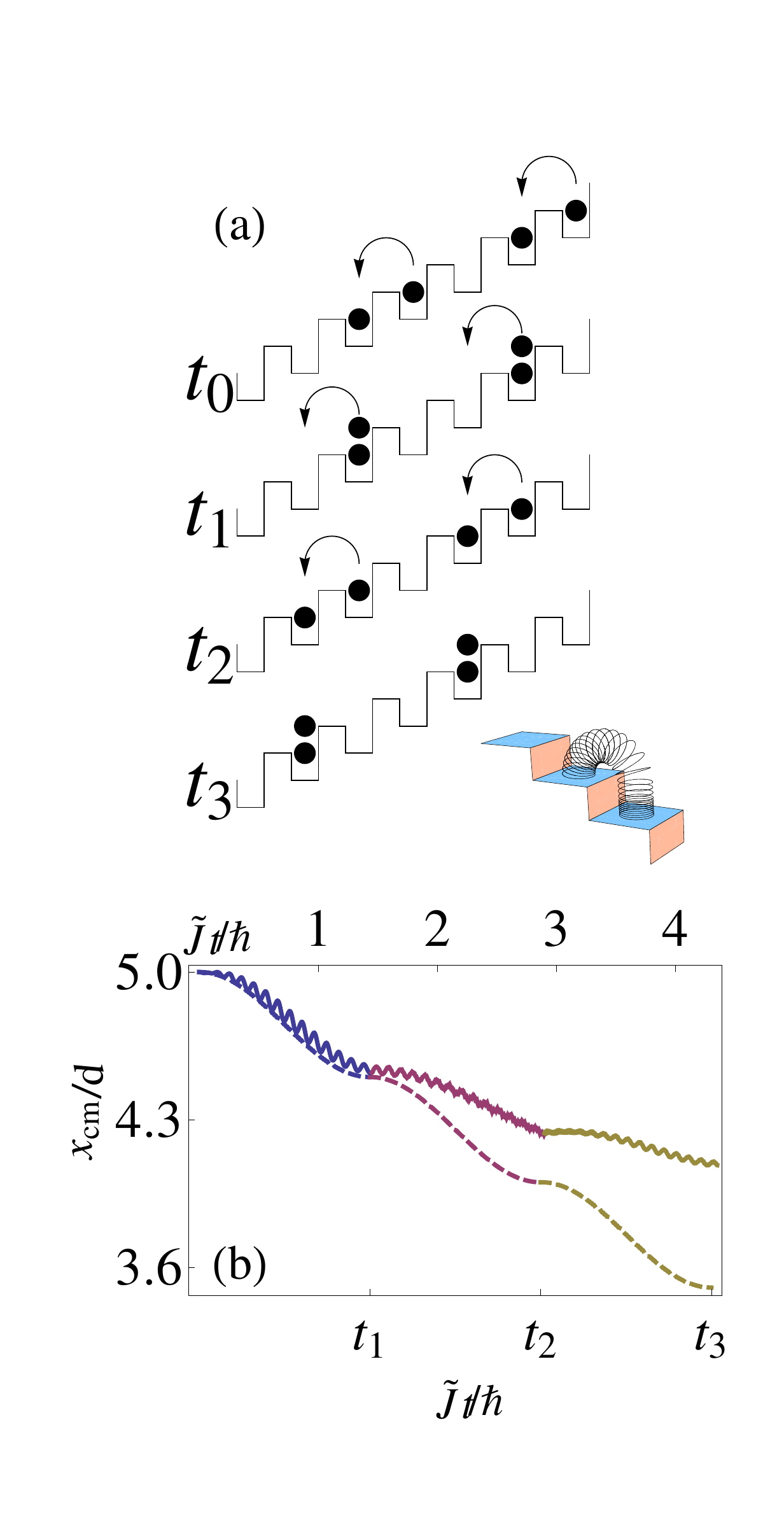} 
\caption{\label{fig:fig10} (Color online) (a) Initial state at $t_0$ obtained by a pattern loading scheme and `slinky motion' obtained by a stroboscopic modulation of the lattice depth with the two resonant frequencies, $\omega = U_0 \pm \Omega_0$.  
The time sequence at which $\omega$ is alternated is: $t_0 = 0$, $\tilde{J} t_1 = \frac{\pi}{\sqrt{2}}$, $t_2 = 2 t_1$, and $t_3 = 3 t_1$.  One can compare the motion of the particles to that of a toy slinky tumbling down a set of stairs as depicted above.  (b) CM motion due to the stroboscopic modulation of the lattice depth.  
The different colors indicate the intervals in which $\omega = U_0 \pm \Omega_0$, respectively. 
The dashed curve displays the dynamics obtained by fully diagonalizing a BHM, Eq.~(\ref{eq:bhm}), with $L=7$ and $N=4$ in the approximation $U(t)=U_0$, $\Omega(t) = \Omega_0$, and $J(t) \approx \tilde{J} \sin(\omega t)$, captured by Eq.~(\ref{eq:22}). 
The solid line displays results for the same system but in which $J(t)$, $U(t)$, and $\Omega(t)$ are  obtained from Wannier orbitals for $V_{0,x} = 0.5$ and $v=0.3$.   
The time sequence in this case is $\tilde{J} t_1 \approx 1.43$, $t_2 = 2 t_1$, and $t_3 = 3 t_1$, where $\tilde{J} t_1$ is identified numerically as the time at which the first minimum in $x_{\rm cm}(t)$ appears. 
The numbers on the top axis refer to the exact dynamics (solid lines). 
}
\end{figure}

In this section, we apply the projection onto resonant families of small cluster states to the problem of enhancing transport of atoms on optical lattices. 
In particular, we treat systems which are prepared so that only decoupled $|11 \rangle$ clusters are present in the initial state. 
This can be realized using a pattern loading scheme in which an optical superlattice is generated by superimposing two lattices with different periodicity. 
Such a spatially selective loading of particles onto an optical lattice has been experimentally achieved, as discussed in Ref.~\onlinecite{Peil:2003p2891}. 

The transport through the system now is achieved by applying time-dependent fields.  Driven tunneling by using time-depedent fields has been addressed at the single-particle level before \cite{Grifoni:1998p229, Klumpp:2007p2299, Gaul:2009p255303, Haller:2010p2877}.
For instance, such an effect has been realized in Ref.~\onlinecite{Haller:2010p2877}, where shaking the lattice, {\it i.e.}, by applying a time-dependent linear field $\Omega(t) = \Omega_0(1+ \gamma \sin(\omega t))$ and 
tuning the frequency close to that of the BO leads to an enhancement of the CM motion, called ``super'' Bloch oscillations.  Similar to the approach of Ref.~\onlinecite{Creffield:2007p2875}, we stroboscopically apply two oscillatory driving fields in order to enhance the amplitude of the oscillations. 
At the many-body level, we propose to enhance transport of the atoms by performing an amplitude modulation along the lattice direction, $V_{x}(t) =V_{0x}(1+ v \sin(\omega t))$ where $v \ll 1$ and $V_x$ the amplitude of the lattice potential. 
The effect of the modulation can be addressed by assuming that the Wannier functions possess a Gaussian profile on each lattice site. 
Using the Gaussian approximation \cite{Jaksch:1998p2881}, the  parameters of the BHM are modified and one obtains \cite{Hild:2006p2890} 
\begin{equation}
\begin{split}
J & = J_0 \, e^{-v \sin\left(\omega t\right)}, \\
U & = U_0 \, \left[1+v \sin\left( \omega t \right) \right]^{1/4}, \\
\Omega & = \Omega_0 \, \left[1 + v \sin \left(\omega t \right) \right].
\end{split}
\end{equation}
The linear approximation of the Hamiltonian around $v = 0$ results in $J(t) \approx J_0 + \tilde{J} \sin\left(\omega t \right)$, where $\tilde{J}=V_{0x}J_0v(\frac{d \text{ln}U}{dV_{x}}|_{v=0}-\frac{d\text{ln}J}{dV_{x}}|_{v=0})$.


In the following, we apply the pattern loading scheme to realize an initial state at time $t_0$ as shown in Fig.~\ref{fig:fig10}(a). 
In this state, we position two adjacent particles on every three lattice sites. 
For $\Omega_0 \gg J$, BO are suppressed so that only resonant motion within the $|11\rangle$ cluster is possible. 
Hence, we can treat the dynamics to a good approximation by projecting onto $|11\rangle$ clusters.  
 In addition, we require $U_0 \gg J$ and $|U_0-\Omega_0| \gg J$. 
Then the two energies $U_0 - \Omega_0$ and $U_0+\Omega_0$ are well separated from each other. 
Only by adjusting the frequency of the modulation to either $U_0 - \Omega_0$ or $U_0 + \Omega_0$, resonance of the $|11\rangle$ state with a particle hole excitation can be achieved; in this way, the $|11\rangle$ clusters can be resonant with either the $|20\rangle$ or the $|02\rangle$ configurations. 
Hence, it is possible to control the direction of the motion of the atoms by tuning the frequency accordingly, 
and the time evolution of the two particles in a cluster is either $|\psi(t)\rangle = c_g(t) |11 \rangle + c_e(t) |20 \rangle$ or $|\psi(t)\rangle = c_g(t) |11 \rangle + c_e(t)|02\rangle$, respectively. 
Note that for a doubly occupied site, the effect is reversed, and applying the same driving frequency will lead to a motion in the direction opposite to the motion induced on a pair of neighboring particles. 
In the following, we explore this to formulate our proposal for a transport scheme. 

In order to provide a more quantitative description of this behavior we first set 
 $J(t) \approx \tilde{J} \sin(\omega t)$.  
 Since we have only $|11\rangle$ clusters we can restrict the dynamics to a two-level system described by the effective spin model of Eq.~(\ref{eq:spinmodel}) which in this case is
\begin{equation}
H_{eff}=-\sqrt{2} \tilde{J} \sin \left( \omega t \right) \sigma_x + \frac{\Delta \omega}{2}\sigma_z, 
\end{equation}
where $\Delta \omega = U_0 \pm \Omega_0$. 
In the rotating wave approximation the Hamiltonian is
\begin{equation}
H_{eff}=\frac{\delta}{2}\sigma_z+\frac{\sqrt{2}}{2}J\sigma_y,
\end{equation}
where we have set $\delta = \Delta \omega - \omega$, and we assume $\delta \ll \Delta \omega$.
In this two-state representation, the center of mass observable is $\hat{x}_{cm}\sim \frac{\sigma_z}{4}$.
Assuming an initial state in which $c_g(0)=1$, we obtain
\begin{equation}
x_{cm}(t) \sim \pm \frac{\tilde{J}^2 \cos\left(t \sqrt{2 \tilde{J}^2+\delta ^2}\right)}{4 \tilde{J}^2+2 \delta ^2},
\label{eq:22}
\end{equation}
with the $\pm$ indicating motion up or down.
Even though the CM amplitude is $2\tilde{J}^2/(4\tilde{J}^2+2\delta^2) \leq 1/2$, it is nevertheless possible to implement transport through 
the lattice by stroboscopically alternating the modulation frequencies
 $\omega = U_0 \pm \Omega_0$ in intervals of $\Delta t =  \pi/\sqrt{2}\tilde{J}$ at zero detuning:
in this way, we transform the initial $|11\rangle$ cluster to a double occupancy, and due to the alternation of the frequency, this is further transformed into a $|11\rangle$ cluster {\it shifted} by one lattice spacing relative to the original cluster.

Alternating $\omega$ 
hence leads to a slinky like motion as depicted in Fig.~\ref{fig:fig10}(a), inducing a net transport of  particles through the system. 
Note that the net transport can also be uphill.
In Fig.~\ref{fig:fig10}(b) we show as a proof of principle results for such a slinky motion. 
We display the exact numerical time evolution for a Bose-Hubbard system of $L=7$ sites and $N=4$ particles when stroboscopically modulating $\omega$ and compare it to the result of an  approximation in which $U$ and $\Omega$ are constant in time and $J(t) \approx \tilde{J} \sin(\omega t)$.  
This approximation leads to a CM motion captured by Eq.~(\ref{eq:22}). 
As can be seen, in both cases the CM motion on the time scale treated is strongly enhanced.
At the end of the time evolution shown, the approximation shows transport by $\Delta x_{\rm cm} \approx 1.5$ lattice spacings, while the exact solution shows $\Delta x_{\rm cm} \approx 0.9$ lattice spacings - note that the usual CM motion is restricted to $\Delta x_{\rm cm} \leq 0.5$ lattice spacings. 
Despite the difference between the exact result and the approximate treatment, 
Fig.~\ref{fig:fig10} shows that the description in terms of the slinky motion compares qualitatively.   
We therefore expect that for larger systems transport through the lattice should be realizable. 
In addition, we expect that for our simple example, the transport can further be enhanced by optimizing the parameters.


As mentioned in Sec.~\ref{sec:accuracy_clusterdynamics}, we have neglected the high-frequency BO which will lead to a dephasing of the slinky motion of the atoms.
This, however, can be controlled by choosing $\Omega$ sufficiently large to dampen the  BO  as discussed in Sec.~\ref{sec:accuracy_clusterdynamics}. 
Nevertheless, the tilt must  be weak enough so that the description of the system by a one-band model remains valid. 


\section{Summary}
\label{sec:summary}

We have investigated the resonant dynamics of strongly-interacting bosonic particles on a one-dimensional tilted optical lattice.
At commensurate fillings when tuning to resonance $U = \Omega$, we find  CM oscillations enhanced
 compared to  the standard BO exhibited by non-interacting atoms.
Following Ref.~\onlinecite{Sachdev:2002p458}, the resonant dynamics can be captured by an effective spin-$1/2$ model.
Interestingly, we find signals  of the critical point in the dynamics of the CM oscillations in the  spin  model which calls for further studies.
We develop a method to describe the dynamics at low fillings based on projections onto  small clusters.
We find that this approach provides a good description of the dynamics  up to fillings $n \approx 2/3$.
Using this approximation, we propose a scheme  to engineer transport in the lattice  by stroboscopically applying amplitude modulated frequencies of the lattice and envisage that this scheme can be realized in ongoing experiments on this system \cite{Simon:2104p2830}.

In future work, it would be interesting to explore extensions of this work to study transport in more complex systems, such as systems with higher dimensionality \cite{Witthaut:2004p41, Trompeter:2006p053903}, or systems utilizing atoms that have internal structure  \cite{Witthaut:2010p033602}.
In two dimensions, {\it e.g.}, with a tilt along either one or both directions, the transport is likely to be quite different to that seen here in one dimension. 
Furthermore, a natural extension will be to extend these results on one dimensional wires to consider the transport through Atomtronic transistors \cite{Pepino:2009p509} that naturally contain three ports (a base, collector, and emitter) and thus possess a more complicated topology. 

\section*{Acknowledgements}
We acknowledge useful discussions with M. Greiner and funding by PIF-NSF (grant No. 0904017). 

\appendix

\section{CM motion in the fermionized regime} 
\label{sec:appendixA}
In this appendix, we derive Eq.~(\ref{eq:xcm}).  We assume a lattice of infinite size centered at $j = 0$; the time evolution of an initial single particle wavefunction $|\psi^q(t=0)\rangle$ then is
\begin{equation}\label{eqn1}
|\psi^q(t)\rangle = \sum_{n=-\infty}^{\infty} |\phi_n\rangle e^{-i n \Omega t} \underbrace{\langle \phi_n|\psi^q(0)\rangle}_{\equiv f^q_n},
\end{equation}
with the Wannier-Stark states $|\phi_n\rangle = \sum_{j=-\infty}^{\infty} J_{j-n}(\alpha)|j\rangle$ and $\alpha = 2J/\Omega$.  With this we obtain for the CM motion
\begin{equation}\begin{split}
x_{\rm cm}^q(t) &= \frac{1}{N} \sum_j j \langle \psi^q(t) | n_j | \psi^q(t) \rangle \\
&= \frac{1}{N} \sum_j \sum_{n,m} f_m^{q*}f_n^q \langle \phi_m | n_j | \phi_n \rangle e^{-i (n-m)\Omega t} \\
& = \frac{1}{N} \sum_{n,m} \sum_j j f_m^{q*}f_n^q J_{j-m}(\alpha)J_{j-n}(\alpha) e^{-i (k-m)\Omega t}.
\end{split}\label{eqn2}\end{equation}

Using the recurrence relation, $J_{n-1}(\alpha) + J_{n+1}(\alpha) = \frac{2n}{\alpha}J_n(\alpha)$, and the completeness relation of Bessel functions, $\sum_j J_{j-n}(\alpha)J_{j-m}(\alpha) = \delta_{n,m}$, we obtain the identity 
\begin{equation}
\delta_{m,n+1} + \delta_{m,n-1} + \frac{2}{\alpha} n \delta_{m,n} 
= \frac{2}{\alpha} \sum_j j J_{j-n}(\alpha)J_{j-m}(\alpha).
\label{eqn3}
\end{equation}
With this, the CM motion of Eq.~(\ref{eqn2}) takes the form
\begin{equation}\begin{split}
&x_{\rm cm}^q(t) \\ 
&= \frac{\alpha}{2N}\sum_{n,m}f_m^{q*} f^q_n \left(\delta_{m,n+1} + \delta_{m,n-1} + \frac{2}{\alpha} n \delta_{m,n}\right)e^{-i (n-m)\Omega t} \\
&=\sum_n \frac{n}{N} |f^q_n|^2 + \frac{J}{\Omega N}\sum_n \left(f_{n+1}^{q*}f^q_{n}e^{i\Omega t} + f_{n-1}^{q*}f^q_n e^{-i\Omega t}\right)\\
& = \sum_n \frac{n}{N} |f^q_n|^2 + \frac{2J}{\Omega N}\sum_n \text{Re}[f_{n+1}^{q*}f^q_{n}e^{i\Omega t}].
\end{split}
\end{equation}
The CM motion for the $N$-body system is obtained by summing over the single particle states $q$,  resulting in Eq.~(\ref{eq:xcm}).  This expression is exact for an infinite system.  
However, for finite systems for $\Omega \gg J$ we find that this provides an excellent approximation even for system sizes as small as $L=10$. 
Hence, the results from directly diagonalizing the single particle Hamiltonian shown in Fig.~\ref{fig:fig1} and the results obtained from Eq.~(\ref{eq:xcm}) are essentially identical.     


\section{Time-dependent perturbation theory for a double well system}
\label{sec:appendixB}

\begin{figure}[ht]
\includegraphics[width=0.48\textwidth]{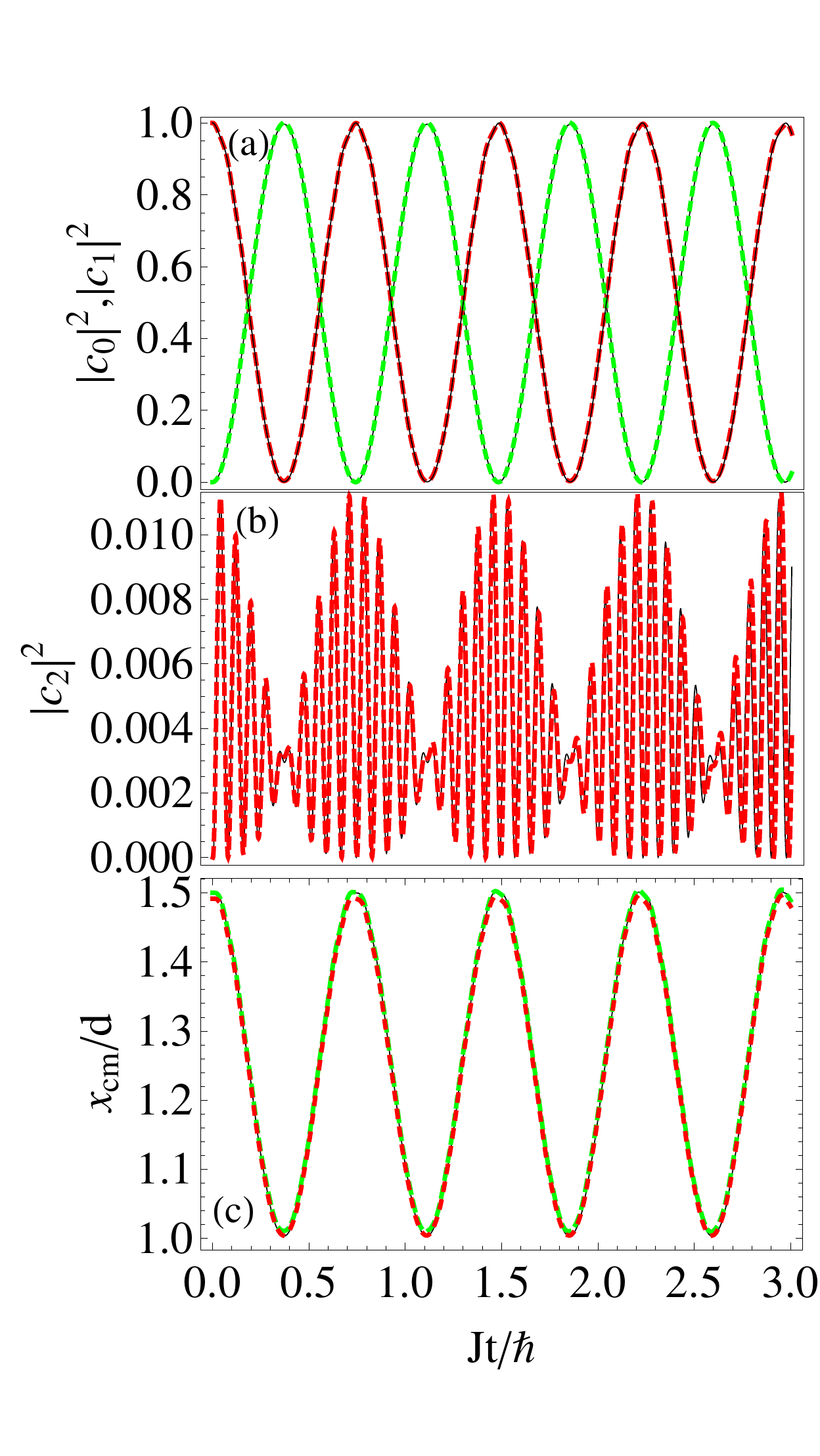}
\caption{\label{fig:fig2appendix} (Color online) 
Coefficients (a) $|c_0(t)|^2, \, |c_1(t)|^2$ and (b) $|c_2(t)|^2$ obtained by solving Eq.~(\ref{eq:DGLscoefficients}) with $J = 3.0$ and $\Omega = 40$. Note that in (b) the magnitude of $|c_2(t)|^2$ is $\sim 1 \%$ of the values shown in (a). The black solid lines depict the exact numerical results whereas the dashed colored lines depict the approximate analytic solutions. The CM motion in (c) compares the exact results [black solid line], results of Eq.~(\ref{eq:b7}) which keeps terms higher than $O(J^2/\Omega^2)$ [green dashes], and the results of Eq.~(\ref{eq:COMdoublewell}) which keeps only terms up to $O(J^2/\Omega^2)$ [red dashes].} 
\end{figure}

The time evolution of the double-well system at resonance and for strong interactions $J \ll \Omega$ is obtained by solving a set of coupled differential equations,
\begin{equation}
\begin{split}
&i \dot{c}_0(t) = - \sqrt{2} \left[J c_1(t) + c_2(t) \right], \\
&i \dot{c}_1(t) = - \sqrt{2} J c_0(t), \\
&i \dot{c}_2(t) = - \sqrt{2} J c_0(t) + 2 \Omega c_2(t),
\end{split}
\label{eq:DGLscoefficients}
\end{equation}
with initial conditions $c_0(0)=1$ and $c_1(0) = c_2(0) = 0$.
The non-resonant state possesses a very small population ($\sim J^2/\Omega^2$) throughout the time evolution [see Fig.~\ref{fig:fig2appendix}(b)] which justifies the zeroth-order approximation $c_2^{(0)}(t)=0$.
The set of equations then simplifies to
\begin{equation}
\begin{split}
&i \dot{c}_0^{(0)}(t) = - \sqrt{2} J c_1^{(0)}(t) \\
&i \dot{c}_1^{(0)}(t) = - \sqrt{2} J c_0^{(0)}(t),
\end{split}
\end{equation}
with the initial condition $c_0^{(0)}(0)=1$.
Hence, $c_0^{(0)}(t)= \cos \left(\sqrt{2}J t \right)$ and $c_1^{(0)}(t)= i \sin \left( \sqrt{2}J t \right)$.  With this, the first order correction on $c_2(t)$ is
\begin{equation}
i \dot{c}_2^{(1)}(t) = - \sqrt{2} J c_0^{(0)}(t) + 2 \Omega c_2^{(1)}(t),
\end{equation}
with the initial condition $c_2^{(1)}(0)=0$.
This gives
\begin{equation}
\begin{split}
&c_2^{(1)}(t) = \frac{e^{-2 i t \Omega } J}{\sqrt{2} \left(J^2-2 \Omega ^2\right)} \\
& \times
 \left[2 \Omega -2 e^{2 i t \Omega } \Omega  \cos\left(\sqrt{2} J t\right) + i \sqrt{2} e^{2 i t \Omega } J \sin\left(\sqrt{2} J t\right)\right].
\end{split} 
\end{equation}
Finally, we address the first order correction to $c_0(t)$ due to $c_2^{(1)}(t)$ by going back to the initial set of equations,
\begin{equation}
i \dot{c}_0^{(1)}(t) = - \sqrt{2} J c_2^{(1)}(t)
\end{equation}
with the initial conditions $c_0^{(1)}(0)=0$.
The solution with the corrections gives
\begin{equation}
\begin{split}
&c_0(t) = \cos\left( \sqrt{2} J t \right)  + \frac{J e^{-2 i t \Omega}}{J^2-2 \Omega ^2}  \\
& \times   \left[ -J + e^{2 i t \Omega } J \cos\left(\sqrt{2} J t\right) - i \sqrt{2} e^{2 i t \Omega } \Omega  \sin\left(\sqrt{2} J t\right)\right]  \\
&c_1(t) = i \sin\left( \sqrt{2} J t \right)  \\
&c_2(t) = \frac{J e^{-2 i t \Omega } }{\sqrt{2} \left(J^2-2 \Omega ^2\right)}  \\
&\times \left[2 \Omega -2 e^{2 i t \Omega } \Omega  \cos \left(\sqrt{2} J t\right) + i \sqrt{2} e^{2 i t \Omega } J \sin\left(\sqrt{2} J t\right) \right] 
\end{split}
\end{equation}
The CM evolution then is
\begin{equation}
\begin{split}
&x_{CM}(t) = \frac{3|c_0(t)|^2+2|c_1(t)|^2+4|c_2(t)|^2}{2} \\
&= \frac{1}{2 \left(J^2-2 \Omega ^2\right)^2}  \left[ 12 J^4-J^2 \Omega ^2 + 10 \Omega ^4 \right. \\
& + \left(3 J^4-7 J^2 \Omega ^2+2 \Omega ^4\right) \cos\left(2 \sqrt{2} J t\right)  \\
& \left. - \, 4 J^2 \left(3 J^2+\Omega ^2\right) \cos\left(\sqrt{2} J t\right) \cos\left(2 t \Omega \right) \right. \\
&\left. -14 \sqrt{2} J^3 \Omega  \sin\left(\sqrt{2} J t\right) \sin\left(2 t \Omega \right) \right].
\end{split}
\label{eq:b7}
\end{equation}
Since we are in the regime $J \ll \Omega$, we neglect terms beyond $J^2/\Omega^2$ to finally obtain Eq.~(\ref{eq:COMdoublewell}),
\begin{equation*}
\begin{split}
&x_{\rm cm}(t) \sim \frac{5}{4}+\frac{\cos\left(2 \sqrt{2} J t\right)}{4} \\
&- \frac{J^2}{ 8\Omega ^2}\left[1+7 \cos\left(2 \sqrt{2} J t\right)+4 \cos\left(\sqrt{2} J t\right) \cos\left(2 t \Omega \right)\right].
\end{split}
\end{equation*}


\section{Coefficients in the cluster dynamics}
\label{sec:appendixC}
We consider a homogeneous system of hard-core bosons at incommensurate filling ($N< L$).  Using open boundary conditions, the single-particle wavefunctions are $\psi_n(x_i) = \sqrt{\frac{2}{L+1}}\text{sin}(\frac{n x_i \pi}{L+1})$ where $n = 1, 2, \dots, L$.  The many-body ground state then is
\begin{equation}
|\psi_0\rangle = \underset{\{\vec{k}\}}{\sum} \, \left| \stackrel{N}{\underset{n,i=1}{{\rm det}}}\psi_n(k_i) \right| \, |\vec{k} \rangle \equiv \underset{\{\vec{k}\}}{\sum} \, c_{\{\vec{k}\}}|\vec{k}\rangle \label{eq:B1}
\end{equation}
 where the sum is over all permutations of the positions of the particles $\{\vec{k}\}$ and the determinant is
\begin{equation}
\stackrel{N}{\underset{n,i=1}{{\rm det}}} \psi_n(k_i)= \left| \begin{array}{cccc}
\psi_1(k_1)& \psi_1(k_2)& \hdots &\psi_1(k_N)  \\
\psi_2(k_1)&\psi_2(k_2) &\hdots & \psi_2(k_N) \\
\vdots & \vdots&\ddots &\vdots   \\ 
\psi_N(k_1) & \psi_N(k_2) &\hdots&\psi_N(k_N) \end{array} \right|.\\
\end{equation}
For a given $\vec{k}$, the determinant of this alternant matrix takes the form
\begin{equation}
\begin{split}
&\stackrel{N}{\underset{n,i=1}{{\rm det}}}\psi_n(k_i) = 
\left(\frac{2}{L+1}\right)^{1/N}(-2)^{N(N-1)/2}  \\
& \times \prod_{i<j}{\left(\cos \frac{k_i \pi}{L+1}-\cos \frac{k_j \pi}{L+1}\right)} 
 \times \prod_{i=1}^{N}\sin \frac{k_i \pi}{L+1}.
\end{split} \label{eq:B2}
\end{equation}
At low filling factors, Eq.~(\ref{eq:B2}) shows that particles tend not to form large clusters, validating the approximation used in Eq.~(\ref{eq:approxgs}). 
The coefficients in Eqs.~(\ref{eq:approxgs}) and~(\ref{eq:bigone}) are then obtained as
\begin{equation}
c_{\{\vec{k}\}} = \left| \stackrel{N}{\underset{n,i=1}{{\rm det}}}\psi_n(k_i) \right| . 
\end{equation}
The coefficient $D^{|n\rangle}_{\vec{k}}$ defined in Sec.~\ref{sec:incommensurate_filling} is obtained numerically by taking a configuration $\vec{k}$ and counting the occurence of clusters $|n\rangle$.  The $n$ particle clusters are identified by their contiguous empty sites.



\begin{thebibliography}{10}

\bibitem{Bloch:2008p943}
I. Bloch, J. Dalibard, and W. Zwerger, Rev. Mod. Phys. {\bf 80},  885  (2008).

\bibitem{Pepino:2009p509}
R.~A. Pepino, J. Cooper, D.~Z. Anderson, and M.~J. Holland, Phys. Rev. Lett.
  {\bf 103},  140405  (2009).

\bibitem{Seaman:2007p2072}
B.~T. Seaman, M. Kr{\"a}mer, D.~Z. Anderson, and M.~J. Holland, Phys. Rev. A
  {\bf 75},  23615  (2007).

\bibitem{Bakr:2009p2641}
W.~S. Bakr, J.~I. Gillen, A. Peng, S. F{\"o}lling, and M. Greiner, Nature {\bf
  462},  74  (2009).

\bibitem{Bakr:2010p1984}
W.~S. Bakr, A. Peng, M.~E. Tai, R. Ma, J. Simon, J.~I. Gillen, S. F{\"o}lling,
  L. Pollet, and M. Greiner, Science {\bf 329},  547  (2010).

\bibitem{Sherson:2010p2701}
J.~F. Sherson, C. Weitenberg, M. Endres, M. Cheneau, I. Bloch, and S. Kuhr,
  Nature {\bf 467},  68  (2010).

\bibitem{Ashcroft:1976}
N.~W. Ashcroft and N.~D. Mermin, {\em Solid State Physics} (Holt, Rinehart, and
  Winston, New York, 1976).

\bibitem{Feldmann:1992p7252}
J. Feldmann, K. Leo, J. Shah, D.~A.~B. Miller, J.~E. Cunningham, T. Meier, G.
  von Plessen, A. Schulze, P. Thomas, and S. Schmitt-Rink, Phys. Rev. B {\bf
  46},  7252  (1992).

\bibitem{Dahan:1996p4508}
M. Ben~Dahan, E. Peik, J. Reichel, Y. Castin, and C. Salomon, Phys. Rev. Lett.
  {\bf 76},  4508  (1996).

\bibitem{Battesti:2004p2768}
R. Battesti, P. Clad{\'e}, S. Guellati-Kh{\'e}lifa, C. Schwob, B. Gr{\'e}maud,
  F. Nez, L. Julien, and F. Biraben, Phys. Rev. Lett. {\bf 92},  253001
  (2004).

\bibitem{Alberti:2009p508}
A. Alberti, V.~V. Ivanov, G.~M. Tino, and G. Ferrari, Nature Physics {\bf 5},
  547  (2009).

\bibitem{Gemelke:2005p2766}
N. Gemelke, E. Sarajlic, Y. Bidel, S. Hong, and S. Chu, Phys. Rev. Lett. {\bf
  95},  170404  (2005).

\bibitem{Ivanov:2008p2767}
V.~V. Ivanov, A. Alberti, M. Schioppo, G. Ferrari, M. Artoni, M.~L. Chiofalo,
  and G.~M. Tino, Phys. Rev. Lett. {\bf 100},  43602  (2008).

\bibitem{Sias:2007p507}
C. Sias, A. Zenesini, H. Lignier, S. Wimberger, D. Ciampini, O. Morsch, and E.
  Arimondo, Phys. Rev. Lett. {\bf 98},  120403  (2007).

\bibitem{Micheli:2004p2862}
A. Micheli, A.~J. Daley, D. Jaksch, and P. Zoller, Phys. Rev. Lett. {\bf 93},
  140408  (2004).

\bibitem{Daley:2005p2859}
A.~J. Daley, S.~R. Clark, D. Jaksch, and P. Zoller, Phys. Rev. A {\bf 72},
  43618  (2005).

\bibitem{Rey:2007p459}
A.~M. Rey, V. Gritsev, I. Bloch, E. Demler, and M.~D. Lukin, Phys. Rev. Lett.
  {\bf 99},  140601  (2007).

\bibitem{Gorshkov:2010p2866}
A.~V. Gorshkov, J. Otterbach, E. Demler, M. Fleischhauer, and M.~D. Lukin,
  Phys. Rev. Lett. {\bf 105},  60502  (2010).

\bibitem{Gluck:2002p103}
M. Gl\"uck, A.~R. Kolovsky, and H.~J. Korsch, Physics Reports {\bf 366},  103
  (2002).

\bibitem{Sachdev:2002p458}
S. Sachdev, K. Sengupta, and S.~M. Girvin, Phys. Rev. B {\bf 66},  75128
  (2002).

\bibitem{Simon:2104p2830}
J. Simon, W.~S. Bakr, R. Ma, M.~E. Tai, P.~M. Preiss, and M. Greiner, Nature
  {\bf 472},  307  (2011).

\bibitem{Pielawa:2011p499}
S. Pielawa, T. Kitagawa, E. Berg, and S. Sachdev, Phys. Rev. B {\bf 83},
  205135  (2011).

\bibitem{Noack:2005p75}
R.~M. Noack and S.~R. Manmana, AIP Conf. Proc. {\bf 789},  93  (2005).

\bibitem{Park:1986p2892}
T.~J. Park and J.~C. Light, The Journal of Chemical Physics {\bf 85},  5870
  (1986).

\bibitem{Hochbruck:1997p2899}
M. Hochbruck and C. Lubich, SIAM Journal on Numerical Analysis {\bf 34},  1911
  (1997).

\bibitem{Manmana:2005p63}
S.~R. Manmana, A. Muramatsu, and R.~M. Noack, AIP Conf. Proc. {\bf 789},  269
  (2005).

\bibitem{white1992}
S.~R. White, Phys. Rev. Lett. {\bf 69},  2863  (1992).

\bibitem{white1993}
S.~R. White, Phys. Rev. B {\bf 48},  10345  (1993).

\bibitem{Schollwock:2005p2117}
U. Schollw{\"o}ck, Rev. Mod. Phys. {\bf 77},  259  (2005).

\bibitem{Daley:2004p2943}
A.~J. Daley, C. Kollath, U. Schollw{\"o}ck, and G. Vidal, Journal of
  Statistical Mechanics: Theory and Experiment {\bf 04},  005  (2004).

\bibitem{White:2004p2941}
S.~R. White and A.~E. Feiguin, Phys. Rev. Lett. {\bf 93},  76401  (2004).

\bibitem{Peil:2003p2891}
S. Peil, J.~V. Porto, B.~L. Tolra, J.~M. Obrecht, B.~E. King, M. Subbotin,
  S.~L. Rolston, and W.~D. Phillips, Phys. Rev. A {\bf 67},  51603  (2003).

\bibitem{Sias:2008p2075}
C. Sias, H. Lignier, Y.~P. Singh, A. Zenesini, D. Ciampini, O. Morsch, and E.
  Arimondo, Phys. Rev. Lett. {\bf 100},  40404  (2008).

\bibitem{Moritz:2003p2981}
H. Moritz, T. St{\"o}ferle, M. K{\"o}hl, and T. Esslinger, Phys. Rev. Lett.
  {\bf 91},  250402  (2003).

\bibitem{Fertig:2005p2980}
C.~D. Fertig, K.~M. O'Hara, J.~H. Huckans, S.~L. Rolston, W.~D. Phillips, and
  J.~V. Porto, Phys. Rev. Lett. {\bf 94},  120403  (2005).

\bibitem{Mun:2007p2985}
J. Mun, P. Medley, G.~K. Campbell, L.~G. Marcassa, D.~E. Pritchard, and W.
  Ketterle, Phys. Rev. Lett. {\bf 99},  150604  (2007).

\bibitem{Strohmaier:2007p947}
N. Strohmaier, Y. Takasu, K. G{\"u}nter, R. J{\"o}rdens, M. K{\"o}hl, H.
  Moritz, and T. Esslinger, Phys. Rev. Lett. {\bf 99},  220601  (2007).

\bibitem{Mckay:2008p2986}
D. McKay, M. White, M. Pasienski, and B. DeMarco, Nature {\bf 453},  76
  (2008).

\bibitem{Ferrari:2006p2792}
G. Ferrari, N. Poli, F. Sorrentino, and G.~M. Tino, Phys. Rev. Lett. {\bf 97},
  60402  (2006).

\bibitem{Carusotto:2005p2828}
I. Carusotto, L. Pitaevskii, S. Stringari, G. Modugno, and M. Inguscio, Phys.
  Rev. Lett. {\bf 95},  93202  (2005).

\bibitem{Wannier:1960p2940}
G.~H. Wannier, Physical Review {\bf 117},  432  (1960).

\bibitem{Hartmann:2004p2847}
T. Hartmann, F. Keck, H.~J. Korsch, and S. Mossmann, New Journal of Physics
  {\bf 6},  2  (2004).

\bibitem{Trombettoni:2001p2848}
A. Trombettoni and A. Smerzi, Phys. Rev. Lett. {\bf 86},  2353  (2001).

\bibitem{Witthaut:2005p2855}
D. Witthaut, M. Werder, S. Mossmann, and H.~J. Korsch, Phys. Rev. E {\bf 71},
  36625  (2005).

\bibitem{Kolovsky:2003p2944}
A.~R. Kolovsky, Phys. Rev. Lett. {\bf 90},  213002  (2003).

\bibitem{Kolovsky:2010p2945}
A.~R. Kolovsky, E.~A. G{\'o}mez, and H.~J. Korsch, Phys. Rev. A {\bf 81},
  25603  (2010).

\bibitem{Gaul:2009p255303}
C. Gaul, R.~P.~A. Lima, E. D\'\i{}az, C.~A. M\"uller, and F.
  Dom\'\i{}nguez-Adame, Phys. Rev. Lett. {\bf 102},  255303  (2009).

\bibitem{Brion:2007p2834}
E. Brion, L.~H. Pedersen, and K. M{\o}lmer, Journal of Physics A: Mathematical
  and Theoretical {\bf 40},  1033  (2007).

\bibitem{Fewell:2005p2841}
M.~P. Fewell, Optics Communications {\bf 253},  125  (2005).

\bibitem{Kolovsky:2004p477}
A.~R. Kolovsky, Phys. Rev. A {\bf 70},  15604  (2004).

\bibitem{Dicke:1954p2982}
R.~H. Dicke, Physical Review {\bf 93},  99  (1954).

\bibitem{brandon_thesis}
B.~M. Peden, Ph.D. thesis, CU Boulder, 2010.

\bibitem{Grifoni:1998p229}
M. Grifoni and P. H\"anggi, Physics Reports {\bf 304},  229   (1998).

\bibitem{Klumpp:2007p2299}
A. Klumpp, D. Witthaut, and H.~J. Korsch, Journal of Physics A: Mathematical
  and Theoretical {\bf 40},  2299  (2007).

\bibitem{Haller:2010p2877}
E. Haller, R. Hart, M.~J. Mark, J.~G. Danzl, L. Reichs{\"o}llner, and H.-C.
  N{\"a}gerl, Phys. Rev. Lett. {\bf 104},  200403  (2010).

\bibitem{Creffield:2007p2875}
C.~E. Creffield, Phys. Rev. Lett. {\bf 99},  110501  (2007).

\bibitem{Jaksch:1998p2881}
D. Jaksch, C. Bruder, J.~I. Cirac, C.~W. Gardiner, and P. Zoller, Phys. Rev.
  Lett. {\bf 81},  3108  (1998).

\bibitem{Hild:2006p2890}
M. Hild, F. Schmitt, and R. Roth, Journal of Physics B: Atomic, Molecular and
  Optical Physics {\bf 39},  4547  (2006).

\bibitem{Witthaut:2004p41}
D. Witthaut, F. Keck, H.~J. Korsch, and S. Mossmann, New Journal of Physics
  {\bf 6},  41  (2004).

\bibitem{Trompeter:2006p053903}
H. Trompeter, W. Krolikowski, D.~N. Neshev, A.~S. Desyatnikov, A.~A.
  Sukhorukov, Y.~S. Kivshar, T. Pertsch, U. Peschel, and F. Lederer, Phys. Rev.
  Lett. {\bf 96},  053903  (2006).

\bibitem{Witthaut:2010p033602}
D. Witthaut, Phys. Rev. A {\bf 82},  033602  (2010).

\end{thebibliography}

\end{document}